\documentclass[12pt]{iopart}
\usepackage{color}
\usepackage{iopams}
\usepackage{graphicx}
\usepackage{mathrsfs}
\usepackage{amsthm}

\newcommand{\bv}{{\mathbf{v}}}

\newcommand{\bS}{{\mathbf{S}}}

\newcommand{\CD}{{\cal D}}
\newcommand{\CE}{{\cal E}}

\newcommand{\CQ}{{\cal Q}}
\newcommand{\CR}{{\cal R}}

\newcommand{\CW}{{\cal W}}
\newcommand{\CM}{{\cal M}}
\newcommand{\average}[1]{\left\langle #1 \right\rangle_\CD}
\newcommand{\ltbaverage}[1]{\left\langle #1 \right\rangle_{LTB}}

\newcommand{\baverage}[1]{\left\langle #1 \right\rangle_{\CD_R}}

\newcommand{\initial}[1]{{#1_{\rm \bf i}}}
\newcommand{\inI}{{\rm I}}
\newcommand{\inII}{{\rm II}}
\newcommand{\inIII}{{\rm III}}
\begin{document}
\review[Toward Physical Cosmology]{Toward physical cosmology:~focus on inhomogeneous\\ geometry and its non--perturbative effects$^{\star}$}
\author{Thomas Buchert}
\address{Universit\'e Lyon 1, Centre de Recherche Astrophysique de Lyon, CNRS UMR 5574\\
9 avenue Charles Andr\'e, F--69230 Saint--Genis--Laval, France\\
Email: buchert@obs.univ-lyon1.fr}
\begin{abstract}
We outline the key--steps toward the construction of a physical, fully relativistic cosmology.  The influence of inhomogeneities on the effective evolution history of the Universe is encoded in backreaction terms and expressed through spatially averaged geometrical invariants. These are absent and potential candidates for the missing dark sources in the standard model. Since they can be interpreted as energies of an emerging scalar field (the morphon), we are in the position to propose a strategy of how phenomenological scalar field models for Dark Energy, Dark Matter and Inflation, that are usually added as fundamental sources to a homogeneous--geometry (FLRW) cosmology, can be potentially traced back to inhomogeneous geometrical properties of space and its embedding into spacetime. We lay down a line of arguments that is -- thus far only qualitatively -- conclusive, and we address open problems of quantitative nature, related to the interpretation of observations.

We discuss within a covariant framework (i) the foliation problem and invariant definitions of backreaction effects; (ii) the background problem and the notion of an effective cosmology; (iii) generalizations of the cosmological principle and generalizations of the cosmological equations; (iv) dark energies as energies of an effective scalar field; (v) the global gravitational instability of the standard model and basins of attraction for effective states; (vi) multiscale cosmological models and volume acceleration; (vii) effective metrics and strategies for effective distance measurements on the light cone, including observational predictions; (viii) examples of non--perturbative models including explicit backreaction models for the LTB solution, extrapolations of the relativistic Lagrangian perturbation theory, and scalar metric inhomogeneities. The role of scalar metric perturbations is critically examined and embedded into the non--perturbative framework.
\end{abstract}
%
\smallskip
%
\bigskip\medskip
\noindent
$^{\star}${\small\it Invited Review for Classical and Quantum Gravity Focus Section: \\``Inhomogeneous Cosmological Models and Averaging in Cosmology''.}
\maketitle

\section{General relativity and cosmology}

\subsection{The foliation issue and the notion of an effective cosmology}

The homogeneous--isotropic standard model of cosmology, being itself a particular solution of Einstein's general theory of relativity, does by far not exploit the degrees of freedom inherent in the geometry as a dynamical variable. It is this richer tone of general relativity -- as compared to the Newtonian theory -- that opens the possibility to generalize cosmological models, notably by including inhomogeneous structure also in the geometrical variables.  There are several guidelines to be emphasized in such a generalization: firstly, a cosmology is thought of as an evolving space section that implies the need to speak of a foliated spacetime, introducing four degrees of freedom (the lapse and shift functions in an ADM setting). This dependence on the foliation should not be confused with the coordinate-- or gauge--dependence of the resulting cosmological equations and variables, however. Secondly, a cosmology 
purports an effective point of view in the sense that the evolving spatially inhomogeneous variables are thought of as being ``averaged over'' in a way that has to be specified.
We aim at a description that only implicitly refers to a metric. However, if a metric is to be specified,
a cosmological metric is then to be considered as an effective, ``smoothed out'' or {\it template metric}, being not necessarily a solution of the equations of general relativity. 
Finally, a {\it physical} cosmology should be characterized by such an effective evolution model, an effective metric to provide the distance scale for the interpretation of  observations, or alternatively an evolution model for average characteristics on the light cone, together with a set of constraint initial data. These latter are to be related to physical properties of fundamental sources, but also to the geometrical data at some initial time (effective, i.e ``averaged'' quantities of known energy sources, but also of intrinsic and extrinsic curvature). This latter clearly emphasizes the absence of any phenomenological parameters. Those would just parametrize our physical ignorance. All these points will be made explicit in what follows.

\subsection{The dark side of the standard model: postulated sources and proposed solutions}

The high level of idealization of the geometrical properties of space in the standard model leads to the need of postulating sources that would generate ``on average'' a strictly, i.e. globally and locally, homogeneous geometry. It is here where a considerable price has to be paid for a model geometry that obviously is not enough to meet physical reality.
Assuming a FLRW geometry $96$ percent of the energy content is missing in the form of a) a postulated source acting attractive like matter, so--called Dark Matter ($\cong 23$ percent) and b) a postulated source acting repulsive, so--called Dark Energy ($\cong 73$ percent). Evidence for the former does indeed come from various scales (galaxy halos, clusters and cosmological, see e.g. \cite{roos}), while evidence for the latter only comes from the apparent magnitude of distant supernovae (see \cite{SNIa:Union,SNIa:Constitution,SNIa:Essence} for the latest data) that, if interpreted within standard model distances, would need an accelerating model. In the simplest case this volume acceration is achieved by a homogeneous--isotropic cosmology with a cosmological constant. 
It should be emphasized that when we speak of evidence, we already approach this evidence with model priors \cite{huntsarkar,seikel:acc,cmbobs}.
Keeping this idealization for the geometry of the cosmological model on all scales, one has to conjecture fundamental fields e.g. in proportion to the missing Dark Matter within small--scale systems also on cosmological scales. The search for these fields is one major research direction in modern cosmology. 

Another huge effort is directed toward a generalization of the underlying theory of gravitation. While this would generalize the geometry of the model, it is not clear why most of these efforts go into a generalization of general relativity and not into the generalization of the cosmological model within general relativity. There are certainly good lines of arguments and various motivations in particle physics and quantum gravity to go beyond the theory of Einstein (for reviews see \cite{DE:review}, \cite{DE:pilar}), but the ``dark problem'' may be first a classical one. 

Looking at generalizations of the standard model \emph{within} general relativity can be identified as a third research direction to which we dedicate our attention here. In light of current efforts it is to be considered conservative, since it does not postulate new fundamental fields and it does not abandon a well--tested theory of gravitation \cite{dressing}, \cite{rasanen:de}, \cite{kolb:backreaction} (for reviews on the physical basis of this third approach see \cite{buchert:jgrg,buchert:review} and \cite{rasanen:acceleration}). 
Among the works in this latter field, research that analyzes spherically symmetric exact solutions has been meanwhile developed to some depth, and has determined the constraints that are necessary to explain Dark Energy, on a postulated observer's position within a large--scale void (see \cite{LTB:review,bolejkoandersson,celerier,voidtest,sussman,sarkar:void,mattsson2}
and references therein, as well as the contributions \cite{bolejkoFOCUS} and \cite{marraFOCUS} in this volume).  

\subsection{Fictitious and physical backgrounds: a more realistic cosmological principle}

Perhaps a reason for not questioning the standard model geometry within general relativity and to go for the search for fundamental fields or for generalizations of the laws of gravitation is the following belief: effectively, i.e. ``on average'', the model geometry has to be {\it homogeneous}, since structures should be ``averaged over''. Then, due to observational facts on large scales (the high degree of isotropy of the Cosmic Microwave Background, if the dipole is completely eliminated due to our proper motion with respect to an idealized exactly isotropic light sphere), and first principle priors (the {\it strong} cosmological principle that requires the universe model to look the same at all places and in all directions), the model's {\it geometry} -- supposed to describe the Universe {\it on average} -- is taken to be {\it locally} isotropic. 

Taking this reasoning at face value we must note the following. The notions of homogeneity and isotropy in the standard model are both too strong to be realistic: firstly, local isotropy implies a model that is locally and globally homogeneous, i.e. despite the conjecture that the homogeneous model describes the inhomogeneous Universe ``on average'', this {\it strict homogeneity} does not account for the fact that any averaging procedure, in one way or another, would introduce a {\it scale--dependence} of the averaged (homogeneous) variables \cite{ellisbuchert}. This scale--dependence, inherent in any physical averages, is suppressed. Even if a large {\it scale of homogeneity} exists (we may call this {\it weak homogeneity principle}), the model is in general scale--dependent on scales below this homogeneity scale \cite{sylos:copernican} (see the contribution by Sylos Labini in this volume \cite{sylosFOCUS}). The same is true for isotropy: while the averaged model may be highly isotropic on large scales, a realistic {\it average distribution} on smaller scales is certainly not (we may call this {\it weak isotropy principle}). Correspondingly, a {\it weak cosmological principle} that applies only to the largest scales would be enough to cover the reality needs while still respecting observational evidence.  

We may summarize the above thoughts by noting that, on large scales, a homogeneous--(almost)isotropic {\it state} does not necessarily correspond to a homogeneous--(almost)isotropic {\it solution} of Einstein's equations. These former states are the averages over fluctuating fields and it is only to be expected that the state coincides with a strictly homogeneous solution in the case of absence of fluctuations. In other words, looking at fluctuations first requires to establish the average distribution. Only then the notion of a {\it background} makes physical sense \cite{kolb:voids,kolb:backgrounds}; see also the contribution by Kolb \cite{kolbFOCUS} in this volume. Current cosmological structure formation models, perturbation theories or N--body simulations, are constructed such that the average vanishes on the background of a homogeneous--isotropic {\it solution} \cite{buchertehlers}.  A such chosen reference background may be a {\it fictitious background}, since it arises by construction rather than derivation. On the contrary, a {\it physical background} is one that corresponds to the average (whose technical implementation has to be specified, and which is nontrivial if tensorial quantities like the geometry have to be ``averaged''). A sound implementation of a physical background will be a statistical background where not only solutions but also the ensembles of solutions are averaged.
Having specified such an averaging procedure, a physical cosmological model may then be defined as an evolution model for the average distribution. Despite these remarks it is of course possible that the homogeneous solution forms at the same time the average. A well--known example is Newtonian cosmology \cite{buchertehlers}. It is also conceivable that a Friedmannian background provides, in some spatial and temporal regimes, a good approximation for the average (compare here the analysis of the stability properties of Friedmannian backgrounds \cite{phasespaceFOCUS} in this volume). 

\section{Refurnishing the cosmological framework}

\subsection{Effective evolution of inhomogeneous universe models}

Taking the point of view of generalizing the cosmological model within general relativity by abandoning the strong cosmological principle (strict homogeneity and isotropy on all scales) and replacing it by the weak cosmological principle (existence of a homogeneity scale and restriction to effective states that are almost isotropic on the scale of homogeneity) leads us to a ``rewriting of the rules'' to build the cosmological model. We shall consider the rules that led to the standard model of cosmology and replace them by their more general counterparts. It follows a basically similar framework that displays, however, a signature of inhomogeneity through the occurence of so--called {\it backreaction terms} and through a manifest {\it scale--dependence}. We shall not introduce new principles or assumptions, apart from the above outlined relaxation of the cosmological principle. We shall restrict ourselves to the simplest case of an {\it irrotational dust model}, except in the last section of this review (for generalizations of the dust model \cite{buchert:dust} with non--constant lapse function see \cite{buchert:fluid}, and for additionally non--vanishing shift see \cite{brown1,brown2,larena,clarkson:perturbations,umeh,veneziano2,marozzi}).

\subsection{The key--steps for generalizing the cosmological framework}

We shall now paraphrase the foundations of the cosmological equations of the standard model and give their generalized counterparts.

\smallskip\noindent
$\bullet$ As in the standard model we introduce a foliation of spacetime into flow--orthogonal hypersurfaces. We generalize the notion of {\it Fundamental Observers} to those that are in free fall also in the general spacetime. Although, as in the standard model, this setup depends on the chosen foliation, we presume that this choice is unique as it prefers the fundamental observers against observers that may be accelerated with respect to the hypersurfaces. A general inhomogeneous hypersurface -- contrary to the homogeneous case -- will, in this setting, unavoidably run into singularities in the course of evolution. This is to be expected in a given range of spatial and temporal scales, since we are treating the matter model as {\it dust}.
This is not a problem of the chosen foliation, but a problem of the matter model that has to be generalized, if small--scale structure formation has to remain regular, and this can be achieved by the inclusion of velocity dispersion and vorticity. 

\medskip\noindent
$\bullet$ As in the standard model we confine ourselves to scalar quantities. We replace, however, the homogeneous quantities by their spatial averages, e.g. the homogeneous density $\varrho_H (t)$ is replaced by $\average{\varrho} (t)$ for the inhomogeneous density $\varrho$ that is volume--averaged over some compact domain $\CD$.
We realize the averaging operation by a mass--preserving Riemannian volume average. In some mathematical disciplines and in statistical averages at one instant of time, it may be more convenient to introduce a volume--preserving averager, but thinking of an averaging domain that is as large as the homogeneity scale we have to preserve mass rather than volume. Furthermore, the average is performed with respect to the above--defined {\it Fundamental Observers}. Spatially averaging a scalar $\Psi (t,X^i)$, as a function of Gaussian coordinates $X^i$ and a synchronizing time $t$, is defined as: 
\begin{equation}
\label{average}
\langle \Psi (t, X^i)\rangle_{\cal D} (t): = 
\frac{1}{V_{\cal D}}\int_{\cal D}  \;\Psi (t, X^i) \;d\mu_g\;\;\;,
\end{equation}
with the Riemannian volume element $d\mu_g := \sqrt{g} d^3 X$, $g:=\det(g_{ij})$, and 
the volume of an arbitrary compact domain, $V_{\cal D}(t) : = \int_{\cal D} \sqrt{g} d^3 X$.
Note that within a more general setup that includes lapse and shift functions, we would have to consider the question whether the locally lapsed time is replaced by a global ``averaged time'' that would involve an average over the lapse function \cite{reiris:averaging}. Here, the dust cosmology is already synchronous, so that this question does not arise. 
Note furthermore, that the building of averages is done in the inhomogeneous geometry. The averages functionally depend on the inhomogeneous metric, but this latter needs not to be specified. We may talk of a {\it kinematical averaging} that does not deform the geometry, i.e. that does not change the physical properties of the inhomogeneous 
spacetime.
For other strategies, see \cite{ellisbuchert}, and references therein, Section~\ref{sec:metrics}, as well as the contribution of Wiltshire \cite{wiltshireFOCUS} in this volume.

\medskip\noindent
$\bullet$ We generalize the kinematical laws of the standard model a) for the volume expansion (the Hamiltonian constraint in the ADM formulation of general relativity) and b) for the volume acceleration (Raychaudhuri's equation in the ADM formulation of general relativity) by dropping the symmetry assumption of local isotropy. The general equations are then volume--averaged,
leading to the following general volume expansion and volume acceleration laws (for a volume scale factor, defined by $a_{\CD}\left(t\right):=\left(V_{\cal D}(t) / V_{\cal D}(t_{i})\right)^{1/3}$; the overdot denotes partial time--derivative, which is the covariant time--derivative here) \cite{buchert:dust}:
\begin{equation}
\fl
3\frac{\ddot{a}_{\CD}}{a_{\CD}}  =  -4\pi G\average{\varrho}+\CQ_{\CD}+\Lambda\;\;\;\;\;;\;\;\;\;\;
3H_{\CD}^{2} + \frac{3 k_\CD}{a_\CD^2} = 8\pi G\average{\varrho} -\frac{1}{2}\CW_\CD - \frac{1}{2}\CQ_{\CD}+\Lambda\;,
\label{averagedequations}
\end{equation}
where $H_{\CD}$ denotes the domain dependent Hubble rate $H_{\CD}=\dot{a}_{\CD} / a_{\CD}=-1/3\average{K}$, $K$ is the trace of the extrinsic curvature $K_{ij}$ of the embedding of the hypersurfaces into the spacetime, and $\Lambda$ the cosmological constant.
The {\it kinematical backreaction} $\CQ_{\CD}$ is composed of averaged extrinsic curvature invariants, while $\CW_\CD$ is an averaged intrinsic curvature invariant 
that describes the deviation of the average of the full (three--dimensional) Ricci scalar curvature $\CR$ from a constant--curvature model,
\begin{equation}
\CQ_{\CD}:= \average{K^2 - K^i_{\;j}K^j_{\;i}} - \frac{2}{3}\average{K}^2 \;\;\;\;;\;\;\;\;\CW_{\CD}: = \average{\CR} - \frac{6 k_\CD}{a_\CD^2}\;.
\label{eq:Def-QW}
\end{equation}
The kinematical backreaction $\CQ_{\CD}$ can also be expressed in terms of kinematical invariants, where the extrinsic curvature is interpreted actively in terms of (minus) the expansion tensor:
\begin{equation}
\CQ_{\CD}:=\frac{2}{3}\left(\average{\theta^{2}}-\average{\theta}^{2}\right)-2\average{\sigma^{2}}\;,
\label{eq:Def-Q}
\end{equation}
where $\theta$ is the local expansion rate and $\sigma^{2}:=1/2 \sigma_{ij}\sigma^{ij}$
is the squared rate of shear. Note that $H_{\CD}$ is now defined as $H_{\CD}=1/3\average{\theta}$.
$\CQ_{\CD}$ appears as a competition term between the 
averaged variance of the local expansion rates, $\average{\theta^{2}}-\average{\theta}^{2}$,
and the averaged square of the shear scalar $\average{\sigma^{2}}$ on the domain
under consideration. 

For a homogeneous domain the above backreaction terms $\CQ_\CD$ and $\CW_\CD$, being covariantly defined and gauge invariants (to second order) in a perturbation theory on a homogeneous background solution, are zero. They encode the departure from homogeneity in a coordinate--independent way \cite{gaugeinv,veneziano2,marozzi}.

The integrability conditions connecting the two Eqs.~(\ref{averagedequations}), assuring that the expansion law is the integral of the acceleration law, read:
\begin{equation}
\langle\varrho{\dot\rangle}_\CD + 3 H_\CD \average{\varrho} \;=0\;\;\;;\;\;\;\;
a_{\CD}^{-2}( a_{\CD}^{2}{\CW}_\CD {\dot )}\;+\;a_{\CD}^{-6}( a_{\CD}^{6}{\CQ}_{\CD} {\dot )}\;=0\;.
\label{eq:integrability}
\end{equation}
While the mass conservation law for the dust is sufficient in the homogeneous case, there is a further equation connecting averaged intrinsic and extrinsic curvature invariants in the inhomogeneous case. The expressions in brackets are conformal invariants (for further details see \cite{buchert:review}).

\subsection{Interpretation and Discussion}

The interpretation of these average equations as {\it generalized or evolving backgrounds} \cite{buchert:review}, \cite{kolb:backgrounds} implies that the new second conservation law in Eq.~(\ref{eq:integrability}) describes an interaction between structure formation and background curvature. In the standard model this latter is absent and structures evolve independently of the background having homogeneous geometry. This homogeneous curvature background furnishes the only solution of (\ref{eq:integrability}), in which structure formation decouples from the background (the expressions in brackets in the second conservation law are separately constant).
Backreaction on such a fixed background decays in proportion to the square of the density and is unimportant in the Late Universe \cite{buchert:dust, buchert:darkenergy, buchert:review}. This  degenerate case of a decoupled evolution explains the fact that in Newtonian and quasi--Newtonian models backreaction has no or little relevance \cite{buchert:review}; in the Newtonian case \cite{buchertehlers}, as well as in Newtonian \cite{bks,abundance} and spatially flat, relativistic spherically symmetric dust solutions (see Section~\ref{sec:backreactionmodels}), $\CQ_\CD$ vanishes. In models with homogeneous geometry and with periodic boundary conditions imposed on the inhomogeneities on some scale, the backreaction term is globally zero and describes cosmic variance of the kinematical properties in the interior of the periodic universe model.

In general, a physical background ``talks'' with the fluctuations, and it is this coupling that gives rise to an instability of the constant--curvature backgrounds as we discuss below.
The essential effect of backreaction models is not a large magnitude of $\CQ_{\CD}$, but a 
dynamical coupling of a nonvanishing $\CQ_{\CD}$ to the averaged scalar curvature deviation $\CW_\CD$. This implies that the temporal
behavior of the averaged curvature deviates from the behavior of a constant--curvature model. In concrete studies, as discussed in Section~\ref{sec:multiscale}, this turns out to be the major effect of backreaction, since it does not only change the kinematical properties of the cosmological model, but also the interpretation of observational data as we explain in Section~\ref{sec:metrics}.

\section{Scalar field language for backreaction: the morphon}
\label{sec:scalarfields}

\subsection{Rewriting the averaged equations as an effective Friedmannian model}

We rewrite the above set of spatially averaged equations together with their integrability conditions by appealing to the kinematical equations of the standard model, which will now be sourced by an {\it effective} perfect fluid energy--momentum tensor \cite{buchert:fluid}:
\begin{eqnarray}
\fl
3\frac{{\ddot{a}}_{\CD}}{a_{\CD}}  =  - 4\pi G(\varrho_{\rm eff}^{\CD}+3{p}_{\rm eff}^{\CD})+\Lambda \;;\;
3H_{\CD}^{2}- \frac{3 k_\CD}{a_\CD^2}= 8\pi G\varrho_{\rm eff}^{\CD}+\Lambda \;;\;
{\dot{\varrho}}_{\rm eff}^{\CD}+3H_{\CD}(\varrho_{\rm eff}^{\CD}+{p}_{\rm eff}^{\CD})=0\,,\nonumber\\
\label{eq:effectivefriedmann}
\end{eqnarray}
where the effective densities are defined as 
\begin{eqnarray}
\varrho_{{\rm eff}}^{{\CD}} := \average{\varrho} + \varrho_{\Phi}\;\;\;;\;\;\varrho_{\Phi}  :=  -\frac{1}{16\pi G}{\CQ}_{{\CD}}-\frac{1}{16\pi G}\CW_\CD \;;\nonumber
\label{eq:equationofstate}\\
{p}_{{\rm eff}}^{{\CD}} := p_{\Phi}\;\;\;;\;\;\,\quad\qquad p_{\Phi} :=  -\frac{1}{16\pi G}{\CQ}_{{\CD}}+\frac{1}{48\pi G}\CW_\CD \;.
\end{eqnarray}
In this form the effective equations suggest themselves to interpret the extra fluctuating sources in terms of a scalar field \cite{buchert:jgrg,buchert:static,morphon}, which refers to the inhomogeneities in geometrical variables.

\subsection{Scalar field emerging from geometrical inhomogeneities}

In making this suggested analogy concrete we, thus, choose to consider the averaged model as a (scale--dependent) ``standard model'' with matter source evolving in a {\it mean field} of backreaction sources.
This mean field we call the {\it morphon field}, since it captures the morphological (integral--geometrical \cite{buchert:review}) signature of structure. (Note that in more general cases, involving lapse and shift functions, the structure of the scalar field theory suggested by the equations may no longer be a minimally coupled one.) We rewrite \cite{morphon}:
\begin{equation}
\label{morphon:field}
\varrho^\CD_{\Phi}=\epsilon \frac{1}{2}{{\dot\Phi}_\CD}^2 + U_\CD\;\;\;;\;\;\;p^\CD_{\Phi} =
\epsilon \frac{1}{2}{{\dot\Phi}_\CD}^2 - U_\CD\;\;,
\end{equation}
where $\epsilon=+1$ for a standard scalar field (with positive kinetic energy), and 
$\epsilon=-1$ for a phantom scalar field (with negative kinetic energy; if $\epsilon$ is negative, a ``ghost'' can formally arise on the level of an
effective scalar field, although the underlying theory does not contain one; note also that there is no violation of energy conditions, since we have only dust matter).
Thus, from the above equations, we obtain the following correspondence that can be employed to change between the languages:
\begin{equation}
\label{correspondence1}
-\frac{1}{8\pi G}{\CQ}_\CD \;=\; \epsilon {\dot\Phi}^2_\CD - U_\CD\;\;\;;\;\;\;
-\frac{1}{8\pi G}\CW_\CD = 3 U_\CD\;\;.
\end{equation} 
Correspondence (\ref{correspondence1}) recasts the integrability conditions (\ref{eq:integrability})
into a (scale--dependent) Klein--Gordon equation for $\Phi_\CD$, and 
${\dot\Phi}_\CD \ne 0$:
\begin{equation}
\label{kleingordon}
{\ddot\Phi}_\CD + 3 H_{\cal D}{\dot\Phi}_\CD + 
\epsilon\frac{\partial}{\partial \Phi_\CD}U(\Phi_\CD , \average{\varrho})\;=\;0\;\;.
\end{equation} 
We appreciate that the deviation of the averaged scalar curvature from a constant--curvature model is directly proportional to the potential energy density of the scalar field.
Averaged universe models obeying this set of equations follow, thus, a Friedmannian kinematics with a fundamental matter source, and an effective scalar field source that reflects the shape of spatial hypersurfaces and the shape of their embedding into spacetime. 
Given the potential in terms of the variables of the averaged system, the evolution of these models is fixed (the governing equations are closed). This also potentially fixes coupling parameters,
since all involved fields can be traced back to the initial value problem of general relativity.

\subsection{Interpretation and Discussion}

The morphon formulation of the backreaction problem offers a good interpretation in terms of energies: a homogeneous model, $\CQ_\CD = 0$ (a necessary and sufficient condition to also drop the scale--dependence, if required on every scale), is characterized by the {\it virial equilibrium condition}:  
\begin{equation}
\label{virialbalance}
\fl
2\, E^\CD_{\rm kin} \,+\, E^\CD_{\rm pot} =\;- \frac{\CQ_\CD V_\CD}{8\pi G}\;\;\;\;,\;\;\;\;\CQ_\CD = 0 \;\;\;\;;\;\;\;\;E^\CD_{\rm kin} = \varepsilon{\dot\Phi}_\CD^2 V_\CD \;\;\;,\;\;\;\;
E^\CD_{\rm pot} = -U_\CD V_\CD \;\;.
\end{equation} 
Deviations from homogeneity, $\CQ_\CD \ne 0$, thus invoke a non--equilibrium dynamics of the morphon in its potential that is dictated by the effective intrinsic curvature of the space in which the fluctuations evolve. The off--equilibrium state can be measured by a Kullback--Leibler distance \cite{entropy1,entropy3}, an entropy functional that arises naturally from the non--commutativity of averaging and the density evolution \cite{entropy1}. It is conjectured \cite{entropy1} and shown to hold in some popular models \cite{entropy2} that this entropy grows in the course of structure formation.

Morphon energies are redistributed and can be assigned to the {\it dark energies}. Dependent on the signs of the backreaction terms (and a sign change may occur in the course of structure formation and by looking at different spatial scales) the morphon can act as a scalar field model for {\it Dark Matter}, a quintessence model for {\it Dark Energy}, or it may even play the role of a {\it classical inflaton} \cite{buchertobadia:inflation}. (For the different interpretations of scalar fields see the review \cite{DE:review}, and for unified views the selection of papers \cite{arbey,matos:unification,sahni,chaplygin}, and for Scalar Dark Matter e.g. \cite{matos:dm1, matos:halos,matos:dm2}).

\section{Global gravitational instability of the standard model backgrounds}
\label{sec:instability}

\subsection{The phase space of exact background states}

The space of possible states of an averaged cosmological model, or the space of {\it physical backgrounds} has one dimension more than the space of possible homogeneous--isotropic solutions in the standard model framework. This can be seen by introducing adimensional ``cosmological parameters''.   
We divide the volume--averaged expansion law 
by the squared volume Hubble functional $H_\CD := {\dot a}_\CD / a_\CD$ introduced before.
Then, the expansion law can be expressed as a sum of adimensional average characteristics:
\begin{equation}
\label{omega}
\Omega_m^{\CD}\;+\;\Omega_{\Lambda}^{\CD}\;+\;\Omega_k^\CD \;+\;\Omega_{\CW}^{\CD}\;+\;
\Omega_{\CQ}^{\CD}\;=\;1\;\;,\quad {\rm with:}
\end{equation}
\begin{equation}
\fl
\Omega_m^{\CD} : = \frac{8\pi G  \langle\varrho\rangle_{\cal D} }{3 H_{\CD}^2}  \;;\;
\Omega_{\Lambda}^{\CD} := \frac{\Lambda}{3 H_{\CD}^2 }\;;\;
\Omega_{k}^{\CD} := - \frac{k_{\initial\CD}}{a_\CD^2 H_{\CD}^2 }\;;\;
\Omega_{\CW}^{\CD} := - \frac{\CW_\CD}{6 H_{\CD}^2 }\;;\;
\Omega_{\CQ}^{\CD} := - \frac{{\CQ}_{\CD}}{6 H_{\CD}^2 } \;.
\end{equation}
Taking for simplicity the constant--curvature parameter and the curvature deviation into a single full curvature parameter, $\Omega_k^\CD + \Omega_\CW^\CD = :\Omega_\CR^\CD$, the generalized model offers a
{\it cosmic quartet} of parameters. Furthermore, if we put $\Lambda =0$, the expansion law defines, for each scale, a two--dimensional phase space of states. A one--dimensional subset of this phase space is formed by backgrounds with Friedmannian dynamics (for illustrations see Figure 3 in \cite{morphon} or Figures 1 and 2 in \cite{buchert:review}, and especially Figures 1 and 2 in \cite{phasespaceFOCUS}). 

We can analyze the fix points and their stability properties in the general dynamical system \cite{morphon}; see the detailed investigation in this volume \cite{phasespaceFOCUS}. Corresponding results for the LTB solutions support these findings \cite{sussman:instability} and \cite{sussman:review}.
The principal outcome of these studies is that the standard zero--curvature model forms a {\it saddle point}; of particular interest are two instability sectors for the standard model, regarded as averaged state:
firstly, perturbed homogeneous states are driven into a sector of highly isotropic, negative curvature and accelerated expanding ``backgrounds'' where backreaction thus mimics Dark Energy behavior over the domain $\CD$; secondly,  perturbed homogeneous states are driven into a sector of highly anisotropic, positive curvature, collapsing and decelerated ``backgrounds'' where backreaction thus mimics Dark Matter behavior over the domain $\CD$. Concrete models, discussed in Section~\ref{sec:multiscale} show that the former happens on large scales, and the latter on the scales of galaxy surveys, and also on smaller scales; (compare here with the phase space orbits depicted in Figure 7 in \cite{multiscale}).

Thus, qualitatively, the instability sectors identified comply with the aim to trace the dark components back to geometrical properties of space, but they also agree with the expected properties of the structure: isotropic, accelerating states on large scales, and highly anisotropic structures on the filamentary distribution of superclusters. Moreover, the curvature properties also meet the expectations (to be inferred from the averaged Hamiltonian constraint): on large scales the Universe is void--dominated and, hence, dominated by negative curvature, while on intermediate scales over--densities are more abundant and are individually characterized by positive curvature. (A more refined classification of instability sectors, associated with the dark sectors of the concordance model, is provided in \cite{phasespaceFOCUS} in this volume.)

\subsection{Dark Energy and Dark Matter hidden in the geometry of space}

The fact that the standard model can be globally unstable in the phase space of averaged states, and the fact that the instability sectors lie in the right corners to explain Dark Energy and Dark Matter behavior, are both strong qualitative arguments to expect that the conservative explanation of the dark energies through morphon energies is valuable. The underlying mechanism is indeed based on the fundamental existence of the relation between geometrical curvatures and sources dictated by Einstein's equations. 

Whether this mechanism is sufficient in a quantitative sense is to date still an open issue. The difficulty to construct quantitative models is to be seen in the need for non--standard tools, for example perturbation theory on a fixed reference background should be replaced by a fluctuation theory on an evolving background that captures the average over the fluctuations.
The question whether perturbations are small can only be answered if we know with respect to which background they are small. Furthermore, since backreaction affects the geometry, it will change the interpretation of observational data, a problem that is intimately related to the generalization of the cosmological model, and to which we shall come 
in Section~\ref{sec:metrics}.

Before, we shall in the next section explain the identified mechanism by discussing some physical properties of structure formation and its relation to the interpretation of geometrical curvature invariants, and how exactly they mimic the dark sources. We here touch on a deeper problem: backreaction effects qualitatively mimic both, Dark Energy and Dark Matter, simultaneously. Whether, on a given domain, or on an ensemble of domains on a given scale, the  morphon mimics Dark Energy or Dark Matter behavior,  changes as a function of time and as a function of scale. Moreover, the small--scale contribution to e.g. a Dark Matter behavior requires more sophisticated relativistic models than the dust model used throughout here (e.g. \cite{dm1,dm2,dm3}). Considering rotation curves of galaxy halos, missing gravitational sources in clusters or missing sources on cosmological scales  always needs different modeling strategies. We try in the following to provide a first step of disentangling Dark Energy and Dark Matter behavior by explicitly constructing an effective multiscale cosmological model.  

\section{Multiscale cosmology: generic volume partitioning of the Universe}
\label{sec:multiscale}

\subsection{A note on the non--local nature of averaging}

Contrary to the standard model, where a homogeneous background is used as a standard of reference for the expansion history of the Universe, a background constructed as the average over fluctuating fields introduces a subtle element: while a homogeneous geometry can be characterized locally, an average is non--local, since it is determined by the inhomogeneities inside, but also outside the averaging domain, reflecting the non--local nature of gravitation. Furthermore, an average incorporates correlations/fluctuations of the local fields, expressed e.g. through the variance between local and averaged quantities. It is this latter which is the key--driver of a repulsive {\it effective pressure} that arises in the averaged models, as we explain now.

\subsection{Structure--emerging volume acceleration}

The simple fact that the local expansion rates differ from their average value on some scale provides the reason why backreaction can produce a volume--accelerating component despite the decelerating nature of the general local acceleration law. 
We here are not talking about an exotic ingredient that produces such a repulsion; it is a basic physical property of a lumpy matter distribution as was already noted in \cite{buchert:jgrg}.

This physical property can easily be understood by comparing the local and the volume--averaged Raychaudhuri equation (for vanishing vorticity and pressure that both would also act accelerating on the local level, but only on small scales) \cite{buchert:dust}:
\begin{equation}
\label{raychaudhuri}
\fl
\quad\dot{\theta} = \Lambda - 4\pi G \varrho + 2{\rm II} - {\rm I}^2 \;\;\;\;;\;\;\;\; \langle\theta\dot\rangle = \Lambda - 4\pi G \average{\varrho} + 2\average{{\rm II}} - 
\average{\rm I}^2\;\;,
\end{equation}
where we defined the principal scalar invariants of the expansion tensor $\Theta_{ij}$, $2{\rm II}:= 2/3 \theta^2 - 2\sigma^2$ and ${\rm I}:=\theta$.
Clearly, by shrinking the domain to a point, both equations agree. However, evaluating the local and averaged invariants,
\begin{eqnarray}
\fl
\quad
& 2{\rm II}- {\rm I}^2 = -\frac{1}{3} \theta^2 -2\sigma^2\;\;\;\;;\nonumber\\
\fl
\quad
& 2\average{\rm II} - \average{\rm I}^2 = 
\frac{2}{3} \average{(\theta - \average{\theta})^2} - 2 \average{(\sigma - \average{\sigma})^2} - \frac{1}{3} \average{\theta}^2 - 2\average{\sigma}^2\;\;,
\end{eqnarray}
gives rise to two additional, positive--definite fluctuation terms, where that for the averaged expansion variance enters with a positive sign.
Thus, the time--derivative of a (on some spatial domain $\CD$) averaged expansion may be positive despite the fact that the time--derivative of the expansion {\em at all points} in $\CD$ is negative. 

In concrete models this variance is the source of a possible large--scale volume--acceleration that would be assigned to Dark Energy in the standard model, while the averaged shear variance mimics an attractive source that would be missing as Dark Matter in the standard model on cosmological scales. Both terms are competing in the backreaction term $\CQ_\CD$. Since the latter depends on scale, it may act in both ways.

\subsection{Scale--dependence made explicit}

We can go one step further and make the scale--dependence explicit by introducing a union of disjoint over--dense regions $\CM$ and a union of disjoint under--dense regions
$\CE$, which both make up the total (homogeneity--scale) region $\CD$. The averaged equations can be split accordingly yielding for the kinematical backreaction \cite{buchertcarfora}:
\begin{equation}
\CQ_{\CD} \; =\;  \lambda_{\CM}\CQ_{\CM}+\left(1-\lambda_{\CM}\right)\CQ_{\CE}
+6\lambda_{\CM}\left(1-\lambda_{\CM}\right)\left(H_{\CM}-H_{\CE}\right)^{2}\;,
\end{equation}
where $\lambda_{\CM}:=\left|\CM\right|/\left|\CD\right|$ denotes the volume--fraction of the over--dense regions compared with the volume of the region $\CD$. In a Gaussian random field this fraction would be $0.5$ and would gradually drop in a typical structure formation scenario that clumps matter into small volumes and that features voids that gradually dominate the volume in the course of structure formation.

Ignoring for simplicity the individual backreaction terms on the partitioned domains, the total backreaction features a positive--definite term that describes the variance between
the different expansion histories of over-- and under--dense regions. It is this term that generates a Dark Energy behavior over the domain $\CD$ \cite{multiscale} (see also \cite{rasanen:peakmodel} for a model by R\"as\"anen, and \cite{wiltshire:clocks,wiltshire:solutions,wiltshire:timescape,wiltshire:obs} for Wiltshire's model that is based on this term only, but includes a phenomenlogical lapse function to account for different histories in $\CM$ and $\CE$ regions that, this latter, we cannot implement in the synchronous foliation of a multiscale dust model; the reader may find more details in the contributions by R\"as\"anen \cite{rasanenFOCUS} and Wiltshire \cite{wiltshireFOCUS} in this volume). If we model non--zero individual backreaction terms by an extrapolation of the leading perturbative mode in second--order perturbation theory \cite{gaugeinv,li:scale}, which also corresponds to the leading order in a Newtonian non--perturbative model \cite{bks}, then we even produce a cosmological constant behavior over the homogeneity scale $\CD$, see Figure~3 in \cite{multiscale}.
In other words, the fact that, physically, over--dense regions tend to be gravitationally bound, i.e. do not partake significantly in the global expansion, together with a volume--dominance of under--dense regions, already produces a large--scale kinematical pressure as a source of volume acceleration. A homogeneous background simply cannot account for this physical property.

\section{Effective metrics and light cone distances}
\label{sec:metrics}

\subsection{Template metrics and effective distances}

For the construction of an effective cosmological evolution model, as outlined above, a metric needs not be specified.  The need for the construction of an effective metric in these models arises, since measured redshifts have to be interpreted in terms of distances along the light cone. Given an explicit, generic and realistic, inhomogeneous metric, the need for the construction of effective metrics does not arise.
Also, if we succeed to understand the evolution of light cone averages in relation to distances, then also here an explicit metric will not be needed \cite{rasanen:light,rasanen:light2}. Work is in progress to construct effective equations on light fronts by surface averaging optical scalars \cite{lightcone}. In this line Gasperini {\it et al.} in a recent paper made a first step toward defining suitable covariant and gauge invariant light cone averages \cite{lightcone:venezia}.

The idea of an effective cosmological metric comes from the ``fitting problem'', that has been particularly emphasized by George Ellis already in the 70's \cite{ellis:average}. The observation was that an inhomogeneous metric does not average out to a homogeneous 
metric that forms a solution of general relativity. Not only the nonlinearity of the theory, but also simple arguments of a non--commutativity \cite{ellisbuchert} between evolution equations and the averaging operation, give rise to the need to find a ``best--fit'', we may call it ``template'' geometry, that inherits homogeneity and (almost--)isotropy on the large scales and, at the same time, incorporates the inhomogeneous structure ``on average'' (for earlier practical implementations of this problem see \cite{ellisstoeger}, \cite{hellaby:volume}, \cite{futamase1}, \cite{kasai93}, \cite{carfora:RG}, \cite{futamase2}; {compare here the introduction to early work on the backreaction problem, to the fitting problem and the discussion on geometrical optics by George Ellis in this volume \cite{ellisFOCUS}).

For the solution of the {\it fitting problem} various strategies have been proposed (see \cite{ellisbuchert} and references therein). One strategy, that allows to explicitly perform a ``smoothing'' of an inhomogeneous metric into a constant--curvature metric at one instant of time, is based on Ricci--flow theory: one notices that a smoothing operation of metrical properties can be put into practice by a {\it rescaling} of the metric in the direction of its Ricci curvature. The scaling equations for realizing this are well--studied, and the rescaling flow results in a constant--curvature metric that carries ``dressed'' cosmological variables \cite{klingon}, \cite{dressing}. These incorporate intrinsic curvature backreaction terms describing the difference to the ``bare'' cosmological parameters as they are obtained through kinematical averaging. 

\subsection{Reinterpretation of observational data}

The standard method of interpreting observations is to construct the light cone $ds^2 = 0$ from the line--element $ds^2 = -dt^2 + g^{\rm hom}_{ij} dX^i dX^j$, where the coefficients $g^{\rm hom}_{ij}$ are given in the form of a constant--curvature (FLRW) metric, and then to calculate the luminosity distance $d_L (z)$ in this metric for a given observed redshift $z$. Assuming this metric for the inhomogeneous Universe implies the conjecture that the FLRW metric is the correct ``template'' of an effective cosmological metric. However, the integrated exact equations (the integral properties of a general inhomogeneous model) are not compatible with this metric, simply because the averaged curvature is assumed to be of the form
$\average{\CR} = 6 k a^{-2}$ on all scales. Improving the metric template slightly, by replacing the global scale factor $a(t)$ through the volume--scale factor $a_\CD (t)$ and the integration constant $k$ through the domain--dependent integration constant $k_\initial\CD$, one renders this metric implicitly scale--dependent \cite{singh2}. As we explained, this is not enough since the averaged curvature couples to the inhomogeneities and in general deviates from the $a_\CD^{-2}$--behavior. What we can do as a first approximation, and this would render the metric compatible with the kinematical average properties, is to introduce the exact averaged curvature in place of the constant curvature in this metric form \cite{morphon:obs}.

The resulting effective spacetime metric consists of a synchronous foliation of constant--curvature metrics that are, however, parametrized by the exact integral properties of the
inhomogeneous curvature, thus they ``repair'' the standard template metric as for the evolution properties of spatial variables. Such a construction can be motivated by Ricci--flow smoothing, that guarantees the existence of smoothed--out constant curvature sections at one instant of time, and by assuming that the intrinsic backreaction terms are subdominant, so that we can parametrize the metric by ``bare'' kinematical averages. To stack these hypersurfaces together introduces, however, an inhomogeneous light cone structure \cite{mersinik(t)}, \cite{rasanenk(t)}. Ideally, one would wish to smooth the light cone too, which is also possible by employing Ricci flow techniques. Improving this first approach to a template metric is needed and this is work in progress.

The result of employing improved template metrics as described above is a change in the luminosity distance. It will, e.g., take care of the fact that light mostly propagates in under--dense regions of negative curvature. This will alter the interpretation of all observational data formerly based on FLRW distances. A comparison of the luminosity distances in the multiscale models investigated in \cite{multiscale}, that are based on the template metric of \cite{morphon:obs}, with a flat $\Lambda$CDM model is presented in Figure~2 of \cite{multiscale}, see also \cite{multiscaleP}. Clarkson's $C$--function \cite{clarkson} (see the review by George Ellis in this volume \cite{ellisFOCUS}) features a clear minimum at redshifts of around $3-5$, which may serve as observational evidence for the effective cosmologies, as proposed in \cite{morphon:obs}. As the investigated multiscale models show, it is not even necessary to measure derivatives of distances, since the feature is already present in the distance, and this at smaller redshifts. 
Although this investigation certainly needs refinement, we already appreciate a signature of the different curvature evolution that furnishes a clearcut prediction for future observations (see \cite{morphon:obs} for details on the construction of template metrics, fits to observational data, and predictions).

\subsection{Strategies for generalizing numerical simulations}

The architecture of Newtonian simulations does not allow to describe the generic interaction between structure formation and the background geometry. This latter is, by the very nature of a Newtonian description within a coordinate frame comoving with a (possibly relativistic) background solution, a foliation of non--dynamical Euclidean space sections. Moreover, kinematical backreaction terms vanish on the simulation box due to the technical restriction to a torus topology, see \cite{buchertehlers} for the proof.

Obviously, these restrictions have to be overcome, if we wish to conduct realizations of a physical cosmological model that go beyond the status of toy realizations to study structure formation. Thus, the need to construct relativistic simulations with dynamical geometry arises. Hereby one would wish that well--developed simulation techniques could be used and eventually interpreted within a relativistic framework.
A straightforward method would be to integrate the ADM equations of general relativity. To date only special general relativistic systems are studied numerically, an effort to construct cosmological simulations awaits an attempt.
In this situation we may ask whether one could improve the architecture of Newtonian simulations to render them relevant in a relativistic setting. A first example has been demonstrated in \cite{multiscale}, where the volume fraction between under--dense and over--dense regions has been used as approximate input into the relativistic framework.
A measurement of fluctuations and the kinematical backreaction term could eventually be drawn from a Newtonian simulation and then, iteratively, taken as input into the average equations for a relativistic physical background. The introduction of comoving coordinates in such an evolving background will alter time--scales and the distance interpretation, but it is certainly a rough approximation and follows the spirit of using global template metrics as described above. 

A more systematic strategy is to first use analytical approximation schemes like the relativistic form of Zel'dovich's approximation, that is employed to initialize N--body codes. Work is in progress at this front and we shall discuss this latter strategy in concrete terms in the next section.

\section{Non--perturbative models for backreaction}
\label{sec:backreactionmodels}

Backreaction models can be constructed on the basis of closure conditions on the averaged equations without specifying a metric (for discussions see \cite{buchert:dust}
and \cite{buchert:static})\footnote{Note that the equations for averaged scalars form an infinite hierarchy of equations. Truncating them implies the need for a closure condition. This happens in many physical systems, also for tensorial hierarchies, e.g. the velocity moment hierarchy of the Vlasov equation \cite{adhesive}. We can easily go down one level of this hierarchy, but always the need for closure conditions will arise, if the problem is restricted to a system of ordinary differential equations.}. Examples include globally static and stationary cosmologies \cite{buchert:static}, scaling solutions \cite{morphon}, \cite{multiscale}, \cite{phasespaceFOCUS}, and specifications to other effective equations of state, e.g. \cite{chaplygin}. These approaches only functionally depend on a metric and are to be considered as a motivated ansatz or as balance conditions imposed on the dynamics in the spirit of the virial theorem, where these conditions are not rooted back to explicit solutions of the inhomogeneous dynamics. It is here where a detailed investigation of inhomogeneities is needed to support or discard certain closure conditions.  We already studied the construction of homogeneous template metrics in the last section, which can be made compatible with global assumptions on the evolution of backreaction. Here we are going to study concrete inhomogeneous metric forms and the resulting backreaction models. We start with some general remarks.

\subsection{Some general notes on relativistic metrics and perturbation theories}

Consider for simplicity a {\it spatial} metric form $\bf g$ with coefficients $g_{ij}$ in an exact (co--tangential) basis ${\bf d}X^i \otimes {\bf d}X^j$. We can write any metric as a quadratic form of deformation one--forms, ${\bf g} = \delta_{ab} \,\boldsymbol{\eta}^a \otimes \boldsymbol{\eta}^b$, i.e. in terms of coefficients, $g_{ij} = \delta_{ab} \eta^a_{\;\,i}\eta^b_{\;\,j}$\footnote{We use indices $i,j,k$ to denote coordinate indices, and the indices $a,b,c$ as counters of e.g. forms.}. Now, such a metric form is {\it flat}, i.e. its Ricci tensor vanishes everywhere, if there exist functions $f^a$, such that the one--forms can be written as exact forms, $\boldsymbol{\eta}^a \equiv {\bf d} f^a$. In other words, if we can find a coordinate transformation $x^i = f^{a\equiv i} (X^j , t)$ that transforms the Euclidean metric coefficients in a new basis,  ${\bf d}x^i \otimes {\bf d}x^j$, $\delta_{ij} dx^i dx^j = \delta_{ab} f^a_{\;\,|i} f^b_{\;\,|j} dX^i dX^j$, with a vertical slash denoting partial spatial derivatives, into the metric coefficients $g_{ij}$, then these latter are just a rewriting of the flat space. Given this remark, any perturbation theory that features metric forms of the integrable form, does not describe relativistic inhomogeneities; metric coefficients of the form $g_{ij} = \delta_{ab} f^a_{\;\,|i} f^b_{\;\,|j}$ describe Newtonian (Lagrangian) perturbations on a flat background space. A truly relativistic perturbation theory deforms the background geometry; in other words, the perturbations live in a perturbed space, not on a reference background. This remark also shows that relativistic perturbation terms that are not related to coordinate artifacts can never contain full divergences, since this latter needs integrable one--form fields. 

In light of these introductory remarks, an inhomogeneous relativistic metric produces curvature. The volume--average of this intrinsic curvature on some domain does not obey a conservation law, as can be explicitly seen in the coupling equation to the fluctuations (\ref{eq:integrability}). In particular, intrinsic curvature does in general not average out to zero; for details on curvature estimates see \cite{buchertcarfora}. This fact in itself shows the existence of a dynamical evolution of an averaged curvature, as soon as structures form \cite{buchert:dust}. On the contrary, standard perturbation theory formulated on a fixed background is constructed such that the averages always vanish on the background, demonstrating the limited nature of results obtained by standard perturbation theory. Here, we identify the crucial difference between a fully relativistic cosmology and a quasi--Newtonian model: for the latter, the only fluctuating fields are the matter fields and for them we have a conservation law that assures that over-- and under--densities compensate each other, even for a nonlinear density distribution, while for the former the curvature is also fluctuating; the argument that applies to the density distribution does not apply to the curvature distribution.

Another perturbative argument aims to justify the validity of the homogeneous geometry, even down to the scales of neutron stars \cite{ishibashi,waldgreen}. As argued in \cite{ishibashi} and by many others, perturbations of the metric remain small with respect to the flat background. While this is true, this does not contradict the existence of a large backreaction effect, since these latter depend on {\it spatial derivatives} of the metric that can be large \cite{estim}, \cite{kolb:voids}, \cite{rasanen:perturbation}. Moreover, the perturbations are considered on a background that does not interact with structure. Perturbations may be small on a different (physical) background, in which case a perturbation may already live in a background with strong curvature (a zero--order effect). 

It is therefore not fruitful to argue against the relevance of backreaction within standard limited schemes, but rather an effort to generalize perturbation theory is needed. 
Efforts to construct a fluctuation theory around a physical average are the subject of current studies. Before such a more general theory can be constructed, it is necessary to first look at some results that hold for exact solutions and for approximation schemes describing truely inhomogeneous deformations as outlined above. We shall therefore discuss some results in the next subsection, obtained for the spherically symmetric LTB solution and for the relativistic Lagrangian perturbation theory.
These results will be presented elsewhere within comprehensive investigations \cite{RZA1,RZA2}.

\subsection{Backreaction for spherically symmetric solutions}

A large number of recent publications on the evaluation of backreaction is based on the spherically symmetric LTB solutions (for reference lists 
see the articles \cite{LTB:review,bolejkoandersson,celerier,voidtest,sussman,sarkar:void,mattsson2,sussman:review}, the book by Bolejko {\it et al.} \cite{bolejkoCUP}, as well as the contributions \cite{bolejkoFOCUS} and \cite{marraFOCUS} in this volume). A comprehensive study of LTB solutions in relation to the backreaction formalism discussed in this review and to the existing literature is provided by Sussman \cite{sussman:review}.
This paragraph just focuses on the special character of this class of solutions and provides some exact results. Before we 
come to the relativistic setting, we briefly recall the situation in the Newtonian theory \cite{buchertehlers}, \cite{bks}, \cite{abundance}.

\subsection*{Newton's Iron Spheres} 

In Newtonian theory the background geometry is fixed and given by a Euclidean vector space, the Newtonian spacetime. Let the spatial domain of averaging $\CD=\CD_R$
be taken as a sphere with radius $R$. The velocity $\bv$ inside $\CD_R$ is only
depending on the distance $r$ to the origin and always parallel to the
radial unit vector $\mathbf{e}_r$, $\bv=v(r)\ {\mathbf{e}}_r$.
By doing this we exclude  rotational velocity  fields. The chosen domain
stays spherical at all times.

The Newtonian velocity gradient is denoted by $(v_{i,j})$, with a comma indicating derivative with respect to a non--rotating Eulerian coordinate system; it
may be characterized by its three principal scalar invariants, the trace $\inI$, the dispersion of its non--diagonal components $\inII$, and the determinant $\inIII$ (for explicit expressions we refer the reader to \cite{ehlersbuchert}, \cite{bks}).
The  averaged first invariant  may be  obtained directly  using Gauss' theorem:
\begin{eqnarray}
\fl\qquad
\baverage{\inI(v_{i,j})} & =
\frac{3}{4\pi R^3}\int_{\CD_R}\rmd^3x\,\nabla\cdot\bv =
\frac{3}{4\pi R^3}\int_{\partial \CD_R}\rmd\bS\cdot v(r){\mathbf{e}}_r 
= 3\frac{v(R)}{R} \;, 
\end{eqnarray}
whereas the averages  of the second and third  invariants require some
basic  calculations. One obtains the relations \cite{bks}:
\begin{equation}
\fl\qquad
\baverage{\inII(v_{i,j})} = \frac{1}{3}\baverage{\inI(v_{i,j})}^2 , \quad;\quad
\baverage{\inIII(v_{i,j})} =  \frac{1}{27}\baverage{\inI(v_{i,j})}^3  \;.
\end{equation}
The first of the above relations implies that the backreaction term vanishes identically, $\CQ_{\CD_R}^{\rm spherical} = 2 \baverage{\inII} - \frac{2}{3} (\baverage{\inI})^2 = 0$,
a result which is in accord with Newton's ``Theorem of the Iron Spheres". Since this relation holds true on every scale, the exact averaged equations (\ref{averagedequations}) reduce to the standard FLRW equations.

\subsection*{Some exact results for LTB solutions} 

The LTB solutions for dust generalize the well--known FLRW solutions for dust: their metric not only depends on the time--coordinate as in the FLRW model, but also on the radial coordinate.
The spherical domain can be seen as a superposition of infinitesimally thick homogeneous shells governed by their own dynamics. In a comoving--synchronous setting (see e.g. \cite{LTB:review} for a demonstration but with different notations) the line--element has the form:
\begin{equation}
ds^2=-dt^2+\frac{R'^2(t,r)}{1+2E(r)}dr^2+R^2(t,r)d\Omega^2 \;,
\end{equation}
$E$ being a free intrinsic curvature function of $r$ satisfying $E(r)>-1/2$; the prime denotes partial differentiation with respect to $r$.\\
In this metric, the scalar parts of the Einstein field equations read:
\begin{equation}
4\pi\rho(t,r)=\frac{M'(r)}{R'(t,r)R^2(t,r)} \quad;\quad
\frac{1}{2}\dot R^2(t,r)-\frac{GM(r)}{R(t,r)}=E(r) \;,
\end{equation}
$M$ being another free function of $r$ related to the radial density profile; the overdot denotes partial time--derivative.
Using the relation between the expansion tensor and the metric tensor in the coordinate form
$\Theta^i_{\;j}: =\frac{1}{2}g^{ik}\dot g_{kj}$,
the averaged scalar invariants of the expansion tensor on a simply--connected LTB--domain can be calculated \cite{RZA2}:
\begin{equation}
\ltbaverage{\inI(\Theta^i_{\;j})}=\frac{4\pi}{V_{LTB}}\int_0^{r_{\cal D}}\frac{\partial_r\left(\dot RR^2\right)}{\sqrt{1+2E}}dr \;;
\end{equation}
\begin{equation}
\ltbaverage{\inII(\Theta^i_{\;j})}=\frac{4\pi}{V_{LTB}}\int_0^{r_{\cal D}}\frac{\partial_r\left(\dot R^2R\right)}{\sqrt{1+2E}}dr \;;
\end{equation}
\begin{equation}
\ltbaverage{\inIII(\Theta^i_{\;j})}=\frac{4\pi}{3V_{LTB}}\int_0^{r_{\cal D}}\frac{\partial_r\left(\dot R^3\right)}{\sqrt{1+2E}}dr \;,
\end{equation}
where the volume is given by
\begin{equation}
 V_{LTB}=\frac{4\pi}{3}\int_0^{r_{\cal{D}}}\frac{\partial_r\left(R^3\right)}{\sqrt{1+2E}}dr \;.
\end{equation}
The deviation of the averaged scalar curvature from a constant--curvature model (\ref{eq:Def-QW}) can also be averaged on a LTB domain:
\begin{equation}
\label{eq:curvature-spherical}
{\cal W}_{LTB}=-\frac{16\pi}{V_{LTB}}\int^{r_{\cal D}}_0\frac{\partial_r\left(ER\right)}{\sqrt{1+2E}}dr-6\frac{k_{\initial\CD}V_{\initial {LTB}}^{2/3}}{V_{LTB}^{2/3}}\ .
\end{equation}
These integrals can be straightforwardly  solved for $E(r)=E_0 = const.$ yielding the following exact results for the averaged scalar curvature,
\begin{equation}
\ltbaverage{\cal R} = - \frac{4(E(r)R)'}{R^2R'}= -\frac{4 E_0}{R^2(r_{\cal D})}\;\;,
\end{equation}
and for the averaged invariants:
\begin{equation}
\fl\quad
\ltbaverage{\inII(\Theta^i_{\;j})} = \displaystyle\frac{1}{3}\ltbaverage{\inI(\Theta^i_{\;j})}^2\;\;;\;\;
\label{eq:spherical-I-III}
\ltbaverage{\inIII(\Theta^i_{\;j})} = \displaystyle\frac{1}{27}\ltbaverage{\inI(\Theta^i_{\;j})}^3\,\;.
\end{equation}
Combining the averaged invariants into the backreaction term ${\cal Q}_{LTB}$, {\it cf.} Eq.~(\ref{eq:Def-QW}), we obtain for a spherically symmetric domain with a strong restriction on the curvature function\footnote{The restriction $E=E_0 = const.$ corresponds to self--similar LTB solutions if we require at the same time that the function $M(r) \propto r$ (R.A. Sussman, {\it priv. comm.}), see \cite{sussman}.}:
\begin{equation}
\label{LTBE}
{\cal Q}_{LTB} = 0\ \;,\ \ {\cal W}_{LTB}=0\;.
\end{equation}
We here generalize to non--flat domains a result obtained in \cite{singh1}. Comparing Eq.~(\ref{eq:curvature-spherical}) and the result for the averaged curvature, one can express $k_{\initial\CD}$ as a function of $E_0$: $k_{\initial\CD}=-2E_0 / R^2(\initial t,r_\CD)$.

The result (\ref{LTBE}) mainly shows that spherically symmetric LTB solutions for a geometry with zero intrinsic curvature are quasi--Newtonian, i.e. they are too special and not useful to access the backreaction problem. Only work on LTB solutions that allow for a non--trivial curvature function $E(r)$ and non--constant curvature geometries are relevant in this context. 

The result (\ref{LTBE}) can be interpreted as a generalization of what people have in mind when they quote Birkhoff's theorem, since here the density distribution is continuous and ${\cal Q}_{LTB} = 0$ in general implies that the scale factor, volume--averaged over a spherically symmetric inhomogeneous distribution, follows the FLRW equations.
Note, however, that it is by far not enough to have a spherically symmetric distribution, since the result (\ref{LTBE}) is very special and we cannot expect a similar theorem to hold in a more realistic situation. This point is important since, for this latter reason, we can expect to learn a lot from LTB solutions concerning the backreaction problem
(see especially the review by Sussman \cite{sussman:review}). 

The above subclass of LTB solutions is contained in a wider class of backreaction models that are based on a relativistic Lagrangian approximation scheme. We are going to give some related results in the next subsection.

\subsection{Backreaction in relativistic Lagrangian perturbation theory}
\label{subsect:RZA}

In this subsection we report on an application of relativistic Lagrangian perturbation theory for the construction of a generic backreaction model. 
The formulation of this theory as well as the formalism used to obtain the results sketched here will be published in forthcoming papers \cite{RZA1,RZA2}.
The idea of the construction of this theory is to employ as a single dynamical variable the spatial deformation one--form fields (Cartan's co--frames), $\boldsymbol{\eta}^a$, cast the full set of Einstein's equations for dust matter into a form that features only this variable, and then setup perturbation and solution schemes for this deformation field. Other fields like the $3-$metric are then functionally expressed through solutions of this perturbation variable. 

We write the $3-$metric as a quadratic form of non--normalized co--frames,
\begin{equation}
g_{ij}=G_{ab}\eta^a_{\ i}\eta^b_{\ j}\quad;\quad  g_{ij}(\initial t)=G_{ij}(\initial t)\ .
\end{equation}
To choose non--normalized frames, that has been suggested by \cite{chandra:blackholes}, bears the advantage that the resulting expressions are similar to their Newtonian counterparts (these latter can be found in \cite{bks}).

The obtained perturbation solutions for the deformation one--forms read to leading order (and with a restriction of initial data that eliminates decaying modes):
\begin{equation}
^{\rm RZA}\eta^a_{\ i}(t,X^k): = a(t)\left(\delta^a_{\ i}+\xi(t)\mathscr{P}^a_{\ i}\right)\ ,
\end{equation}
where $\mathscr{P}^a_{\ i}=P^a_{\ i}(\initial t,X^k)$, $\xi(\initial t)=0$, $a(\initial t)=1$.
The function $\xi (t)$ solves the well--known first--order equations to be found in \cite{buchert89,buchert92,bbk}; RZA stands for ``Relativistic Zel'dovich approximation'', generalizing Zel'dovich's idea \cite{zeldovich} and suggested first by Kasai \cite{kasai95} (for normalized co--frames). Contrary to Kasai's definition we consider the full $3-$metric from first--order deformations:
\begin{equation}
\fl
\nonumber^{\rm RZA} g_{ij}(t,X^k)=a^2(t)\left\{G_{ij}+\xi(t)\left(G_{aj}\mathscr{P}^a_{\ i}+G_{ib}\mathscr{P}^b_{\ j}\right)+\xi^2(t)G_{ab}\mathscr{P}^a_{\ i}\mathscr{P}^b_{\ j}\right\}\ .
\end{equation}
One then obtains for the RZA backreaction model in a non--normal basis:
\begin{eqnarray}
\label{resultQ2}
&^{\rm RZA}{\cal Q}_{\cal D}\;=
&\displaystyle\frac{\dot\xi^2\left(\gamma_1+\xi\gamma_2+\xi^2\gamma_3\right)}{\left(1+\xi\langle{\rm I}_{\rm \bf i}\rangle_{{\cal C_D}}+\xi^2\langle{\rm II}_{\rm \bf i}\rangle_{{\cal C_D}}+\xi^3\langle{\rm III}_{\rm \bf i}\rangle_{{\cal C_D}}\right)^2}\;\;,
\end{eqnarray}
where we have defined the set of initial data featuring the initial principal scalar invariants of the expansion tensor (the first is the initial backreaction term):
\begin{eqnarray}
\gamma_1: = 2\langle{\rm II}_{\rm \bf i}\rangle_{{\cal C_D}}-\frac{2}{3}\langle{\rm I}_{\rm \bf i}\rangle_{{\cal C_D}}^2 = \CQ^{\rm initial}_{{\cal C}_D}\;;\nonumber\\
\gamma_2: = 6\langle{\rm III}_{\rm \bf i}\rangle_{{\cal C_D}}-\frac{2}{3}\langle{\rm II}_{\rm \bf i}\rangle_{{\cal C_D}}\langle{\rm I}_{\rm \bf i}\rangle_{{\cal C_D}} \;;\nonumber\\
\gamma_3: = 2\langle{\rm I}_{\rm \bf i}\rangle_{{\cal C_D}}\langle{\rm III}_{\rm \bf i}\rangle_{{\cal C_D}}-\frac{2}{3}\langle{\rm II}_{\rm \bf i}\rangle_{{\cal C_D}}^2 \;.
\end{eqnarray}
We note that for non--normalized co--frames the initial $3-$metric tensor does not appear explicitly in the expression for the backreaction model; the domain $\CD$ is, however, Lagrangian,  i.e. it is frozen into the evolving metric. Initially, the domain $\CD$ can here be chosen to be a section of a Euclidean space, denoted by ${\cal C}_D$. 
All relativistic expressions have a straightforward Newtonian limit by sending the deformation one--forms to exact forms $\boldsymbol{\eta}^a \rightarrow {\bf d} f^{a \equiv i}$, 
where the counting index for the forms $a$ becomes a coordinate index $i$. Note that the Newtonian approximation contains integrable averaged invariants. 

In the situation of a spherically symmetric model restricted to the class of LTB solutions with trivial curvature function $E(r)=E_0 = const.$, the averaged invariants obey the (quasi--Newtonian) relations (\ref{eq:spherical-I-III}), compatible with the backreaction model (\ref{resultQ2}). The generic model (\ref{resultQ2}) therefore contains this class of LTB solutions in a subclass. 
Furthermore, the leading term in the model (\ref{resultQ2}) agrees with the backreaction model of the linear perturbation mode, 
\begin{equation}
\label{linear}
^{\rm RZA}{\cal Q}^{\rm linear}_{\CD} = {\dot\xi}^2 \CQ^{\rm initial}_{{\cal C}_D} = \frac{\CQ^{\rm initial}_{{\cal C}_D}}{a}\;,
\end{equation} 
where the latter equality holds for an Einstein--de Sitter background. This mode plays an important role in the evaluation of the backreaction effect, since it forms the weak--backreaction limit of an exact scaling solution \cite{morphon}:
\begin{equation}
\label{scaling}
^{\rm scaling}{\CQ}^{\rm}_{\CD} = \frac{\CQ^{\rm initial}_{{\cal C}_D}}{a_\CD}\quad;\quad ^{\rm scaling}{\CW}^{\rm}_{\CD} = \frac{\CW^{\rm initial}_{{\cal C}_D}}{a_\CD}\;,
\end{equation} 
that has to be compared with the competing sources in the balance equation of the averaged Hamiltonian constraint (the second of Eqs.~(\ref{averagedequations}))\footnote{For some further remarks in this context see \cite{buchert:review}, Sect. 4.2.},
\begin{equation}
\average{\varrho} \propto \frac{1}{a_\CD^3}\quad;\quad k_\CD \propto \frac{1}{a_{\CD}^2}\quad;\quad ^{\rm scaling}\CQ_\CD \propto \frac{1}{a_\CD}\quad;\quad\Lambda = const \;.
\end{equation}
In Section~\ref{sec:multiscale} we discussed a model that assumes the scaling laws (\ref{scaling}) on subdomains of a multiscale cosmology where the global evolution mimics a cosmological constant behavior as a result of the expansion variance between the subdomains.


\subsection{Backreaction models for relativistic scalar metric inhomogeneities}

Thus far we worked with the matter model `irrotational dust'. To discuss more general matter models such as radiation or the fluid picture of a scalar field, we have to briefly recall
a covariantly defined set of averaged ADM equations obtained previously \cite{buchert:fluid}. These equations are valid for any spacetime foliation within
the class of foliations with vanishing shift, and for any choice of the lapse function and the inhomogeneous 3--metric.
We shall then specify the 3--metric, investigate scalar 
metric inhomogeneities for any choice of the lapse function, and evaluate the relevant 
backreaction terms. This we can do in general without 
resorting to any other approximations than those implied by the restriction to scalars and irrotational flows. We shall also discuss the evolution of backreaction in the longitudinal gauge,
and we shall put forward crucial arguments in favor of a non--perturbative versus a perturbative interpretation of backreaction using this example.

\subsubsection{The averaged ADM equations for vanishing shift}

In this subsection we entirely follow the notations and results given in \cite{buchert:fluid}.
We shall study spatial averages in a hypersurface defined by the choice of the 
in general inhomogeneous lapse function $N (X^i ,t)$ and inhomogeneous $3-$metric coefficients $g_{ij} (X^i ,t)$ in the line--element
\begin{equation}
ds^2 = -N^2dt^2 + g_{ij}dX^idX^j \;\;.
\end{equation}
($X^i$ are local coordinates in a $t=const.$ hypersurface).
This line--element is sufficient to analyze the example of the so--called
longitudinal gauge that we shall consider below. 

Perfect fluid sources are characterized by a diagonal energy--momentum tensor with 
energy density $\varepsilon$ and pressure $p$,
$T_{\mu\nu} = \varepsilon u_{\mu}u_{\nu} + p h_{\mu\nu}$. We may choose to project
onto the fluid's restframe, defining the projection tensor through 
$h_{\mu\nu} = g_{\mu\nu} + u_{\mu}u_{\nu}$,  i.e. we project onto hypersurfaces orthogonal to the fluid's $4$--velocity $u^{\mu}$.
We employ the $4$--velocity of the flow in the form 
\begin{equation}
\label{4velocity}
u^{\mu} = -\frac{\partial^{\mu}{\cal S}}{h}\;\;;\;\;h=\frac{\varepsilon +p}{\varrho}\;\;,
\end{equation}
together with the decomposition into kinematical parts of the $4$--velocity gradient,
\begin{equation}
\label{decomposition}
u_{\mu;\nu} = \frac{1}{3}\theta h_{\mu\nu} + \sigma_{\mu\nu}  + \omega_{\mu\nu}
- {\dot u}_{\mu}u_{\nu}\;\;,
\end{equation}
where the inhomogeneous normalization of the $4$--velocity gradient $h$ 
is given by the injection energy per fluid element and unit restmass,
$d\varepsilon = h d\varrho$ with the restmass density $\varrho$ \cite{israel};
$\theta$ is the rate of expansion, $\sigma_{\mu\nu}$ and $\omega_{\mu\nu}$ the 
shear and vorticity tensors, respectively.

The existence of a scalar $4$--velocity potential $\cal S$ together with the choice
(\ref{4velocity}) implies that the conservation equations $T^{\mu\nu}_{\;\;\,\,;\nu}=0$
are satisfied, but also that the flow has to be irrotational and that the covariant
spatial gradient of $\cal S$ (denoted by a double vertical slash in this paper) vanishes
\cite{brunietal1,brunietal2,dunsbyetal}, \cite{buchert:fluid}:
\begin{equation}
\label{restrictions}
\fl
\omega_{\mu\nu} = h_{\mu}^{\;\,\alpha}h_{\nu}^{\;\,\beta}u_{[\alpha;\beta]} =
- h_{\mu}^{\;\,\alpha}h_{\nu}^{\;\,\beta}\left(\,\partial_{[\alpha}{\cal S}/h\,\right)_{;\beta]}
=0\;\;;\;\;
{\cal S}_{||\mu} = h_{\mu}^{\;\,\alpha}\partial_{\alpha}{\cal S} = \partial_{\mu}{\cal S}
+ u_{\mu} \dot{\cal S} = 0\;,
\end{equation} 
with the covariant time--derivative $\dot {\cal S}:= u^{\mu}{\cal S}_{\,;\mu}\equiv h$.
For the special case of an equation of state of the form $p = \gamma \varepsilon$ we
obtain:
\begin{equation}
\label{gamma}
\fl
\varepsilon = \frac{1}{2\gamma} h^{1 + 1/\gamma}\;\;;\;\;p = \frac{1}{2} 
h^{1 + 1/\gamma}\;\;;\;\;
\varrho = \frac{1+\gamma}{2\gamma} h^{1/\gamma}\;\;;\;\;
{\dot h}+\gamma\theta h = 0\;\;;\;\;h\;\equiv\; \dot{\cal S}\;\;,
\end{equation}
which reduce to the familiar expressions for a free minimally coupled scalar field source
(a ``stiff fluid'' with $\gamma = 1$; this case has been exploited in \cite{buchertveneziano}).

The averaging operation in terms of Riemannian volume
integration is performed, as in the dust case, over the hypersurfaces orthogonal to $u^{\mu}$, 
restricting again attention to the scalar functions $\Psi (X^i ,t)$, {\it cf.} Eq.~(\ref{average}).
We also consider the same definition as for the dust case of a dimensionless volume scale factor
$a_{\cal D} (t)$, which implies that we are only interested in the volume dynamics of the domain; 
$a_{\cal D}$ is a functional of the domain's shape (dictated by the metric) and position. 
As in the dust case we require the domains to follow the flow lines, so that the total restmass 
$M_{\cal D}: = \int_{\cal D} \varrho J d^3 X $
contained in a given domain is conserved. With a non--constant lapse function  
we have, however, to introduce a scaled (t--)expansion ${\tilde\theta}:=N\theta$, 
which describes the rate of change of the domain's volume expansion in the spatial hypersurfaces,
that on average defines an effective Hubble--functional\footnote{We shall reserve the overdot for the covariant time--derivative
(defined through the 4--velocity $u^{\mu}$): $\frac{\partial}{\partial\tau}:= u^{\mu}\frac{\partial}{\partial \mu}=
\frac{1}{N}\frac{\partial}{\partial t}$,
and we abbreviate the coordinate time--derivative by a prime in the sequel.}: 
\begin{equation}
\langle \tilde\theta \rangle_{\cal D} = {\partial_t V_{\cal D} (t) \over 
V_{\cal D} (t)} = 3 
\frac{\partial_t a_{\cal D}}{a_{\cal D}} = 3\frac{a_{\cal D}'}{a_{\cal D}} =: 3 {H}_{\cal D}\;\;\;.
\end{equation}
For an arbitrary scalar field $\Upsilon (X^i ,t)$ we make use of the non--commutativity
relation:
\begin{equation}
\label{commutation}
\fl
\langle \Upsilon\rangle_{\cal D}' - \langle{\Upsilon}'
\rangle_{\cal D} = \langle \Upsilon\tilde\theta\rangle_{\cal D} - 
\langle \Upsilon\rangle_{\cal D}\langle\tilde\theta\rangle_{\cal D}
\,,{\rm or, alternatively},\,
\langle\Upsilon\rangle_{\cal D}' 
+ 3{H}_{\cal D}\langle \Upsilon\rangle_{\cal D}
= \langle\Upsilon' + \Upsilon{\tilde{\theta}}\rangle_{\cal D} .
\end{equation}

Averaging Raychaudhuri's equation and 
the Hamiltonian constraint, we can cast the resulting equations into a compact form
(to be found in {\it Corollary 2} in \cite{buchert:fluid}):
\begin{equation}
\label{effectiveequations}
\fl
3\frac{a_{\cal D}''}{a_{\cal D}} + 
4\pi G \left(\varepsilon_{\rm eff} + 
3p_{\rm eff}\right)\;=\; 0\;;\;
6 {H}_{\cal D}^2 - 16\pi G \varepsilon_{\rm eff} 
\;=\; 0\;;\;
\varepsilon_{\rm eff}' + 3 {H}_{\cal D}
\left(\varepsilon_{\rm eff} + p_{\rm eff} \right)= 0\,,
\end{equation}
with the following fluctuating sources:
\begin{eqnarray}
\label{backreactionsources}
16\pi G \varepsilon_{\rm eff}:= &16\pi G\langle{\tilde\varepsilon}\rangle_{\cal D} 
- {\tilde{\cal Q}}_{\cal D}
- \langle{\tilde{\cal R}}\rangle_{\cal D}\;\;,\qquad\,\;\quad\\
16\pi G p_{\rm eff}:= &16\pi G\langle {\tilde p} \rangle_{\cal D} - {\tilde{\cal Q}}_{\cal D} 
+ \frac{1}{3} \langle{\tilde{\cal R}}\rangle_{\cal D} - 
\frac{4}{3}{\tilde{\cal P}}_{\cal D}\;;
\end{eqnarray}
${\tilde\varepsilon}  := N^2 \varepsilon$ and ${\tilde p}: = N^2 p$ are  
the scaled energy density and pressure of matter, respectively. The 
{\it kinematical backreaction} term is given by:
\begin{equation} 
\label{backreactionQ}
{\tilde{\cal Q}}_{\cal D}:= 2 \langle N^2 II\rangle_{\cal D} 
 - \frac{2}{3} \langle N\theta \rangle^2_{\cal D}\;\;; 
\end{equation}
it is built from the principal scalar invariants 
$2 II: = \theta^2 -K^i_{\,\;j}K^j_{\,\;i}$ 
and $K^i_{\,\;i} = - \theta$ of the extrinsic curvature,
$K^i_{\,\;j} = -\frac{1}{2}{g}^{ik}\frac{1}{N}{g}'_{kj}$.
The averaged $3-$Ricci scalar curvature $\cal R$  and the acceleration terms ({\it dynamical backreaction}) read:
\begin{equation} 
\label{backreactionRandP}
\langle{\tilde{\cal R}}\rangle_{\cal D}:= \langle N^2{\cal R}\rangle_{\cal D}\;\;;\;\;
{\tilde{\cal P}}_{\cal D}:= \langle {\tilde{\cal A}}\rangle_{\cal D}
+ \Bigl\langle \frac{N'}{N}{\tilde\theta}\Bigr\rangle_{\cal D} \;\;,
\end{equation}
with the scaled ($t-$)acceleration divergence ${\tilde{\cal A}}:=N^2 {\cal A} 
= N N^{|i}_{\;\,||i}$\footnote{A single slash denotes partial differentiation
with respect to the coordinates $X^i$, and a double vertical slash covariant
spatial differentiation with respect to the 3--metric as before.}.

\subsubsection{Scalar metric inhomogeneities in a metric form corresponding to the conformal Newtonian gauge}

The so--called conformal Newtonian or longitudinal gauge
is often employed in the study of perturbations on a Friedmannian background cosmology,
and is considered a preferred frame because it offers a well--defined Newtonian limit \cite{wald}\footnote{Note that the framework discussed in
Subsection~7.3 also offers a well--defined Newtonian limit. However, there one falls on the Lagrangian form of the Newtonian equations 
\cite{ehlers,ehlersbuchert:weyl}.}. 
The topic addressed in this review, i.e. the impact of 
inhomogeneities on expansion properties of the Universe (backreaction) is also
often discussed in this gauge \cite{futamase1}, \cite{futamase2}, \cite{abramo1,abramo2},  \cite{branden1,branden2,branden3}, 
\cite{sasaki1}, and many others. In the recent paper \cite{marozzi:inflation} it is shown, in an
inflationary model, how the observers in this gauge are related to the free--falling ones. 
Even  a `no--go conjecture'  has been raised on the issue of whether backreaction can be significant \cite{ishibashi}, 
also advocating the post--Newtonian metric as a sound model for most 
cosmological studies. George Ellis gives a related discussion in this volume \cite{ellisFOCUS}.

Since the averaged equations briefly reviewed above are valid for 
any choice of the lapse function and any ansatz for the $3-$metric, we are in the position to calculate backreaction effects with some generality, i.e.
with no need to invoke approximations other than those implied by the restriction to scalars and irrotational flows. We shall do this calculation explicitly with the aim to illustrate the strict, non--perturbative application of the post--Newtonian metric
form as a solution of general relativity, and to learn some issues about common perturbative interpretations of this metric form.

In the longitudinal gauge the lapse function and the $3-$metric are specified in many studies as to provide
a ``Newtonianly perturbed model'' in the following form:
\begin{equation}
\label{postnewtonlapseandmetric}
N^2 = 1 + 2\phi \;\;\;;\;\;\;g_{ij} = a^2  \left(1 - 2\psi \right)\gamma_{ij}\;\;,
\end{equation}
with the scale--factor $a$  of a homogeneous--isotropic background model, and a
constant--curvature $3-$metric $\gamma_{ij}$. For simplicity, we are going to choose
the Euclidean metric $\gamma_{ij} = \delta_{ij}$ in what follows.
From what has been said previously we are in the position to evaluate all the variables, 
in particular the backreaction terms, as functionals of the lapse function $N (X^i ,t)$, the metrical inhomogeneities $\psi (X^i ,t)$, and their time and space 
derivatives. 

Note already the subtle element that in general relativity $X^i$ are to be local coordinates in a perturbed space. In turn, factoring out a scale factor as in the metric form (\ref{postnewtonlapseandmetric}) implies that, only if $\phi = \psi = 0$, the scale factor obeys the standard Friedmann equations with respect to the coordinate--time; in the perturbed space the scale factor acquires a dependency on $X^i$ as seen in the hypersurfaces $t=const.$, defined by the inhomogeneous lapse function. In the presence of perturbations it makes no physical sense to factor out a function of $t$ that obeys Friedmann's equations. If the background is not perturbed, then
the lapse function can only be time--dependent (for the background equations in hypersurfaces with a time--dependent lapse see
\cite{buchert:fluid}, Sect.~4.1). 

Note also that the simple ansatz for the (conformally flat, i.e. vanishing Cotton--York tensor) metric requires that a scalar function
models all six metric components. It is therefore expected that in the case of a strict application of this metric form as a solution to general relativity we are dealing with a highly restricted situation. We remark that this metric ansatz is general for the case
where the trace--free symmetric part of the extrinsic curvature
(the shear tensor $\sigma_{ij}$) vanishes:
\begin{equation} 
\label{shearfree}
K^i_{\;\,j} = \frac{1}{N}\left[\,\frac{\psi'}{1- 2\psi} - \frac{a'}{a}\,\right]\,\delta^i_{\;\,j} \;\Rightarrow
\;K^i_{\;\,j}-\frac{1}{3}K^k_{\;\,k}\delta^i_{\;\,j} =: -\sigma^i_{\;\,j} = 0\;.
\end{equation}
However, the post--Newtonian form is employed with the implicit 
understanding that $|\phi |$ and $|\psi|$ are small compared to $1$ (together with corresponding
requirements on their derivatives \cite{wald}). For perfect fluids and small peculiar--velocities $\phi = \psi$.
The so--constructed model is designed to be 
in a ``near--Friedmannian state''. In what follows we have to keep in mind that Eq.~(\ref{shearfree}) implies
that any application of this metric--form to describe inhomogeneities can only be considered in an approximate sense. Especially for a non--tilted slicing where 
the expansion tensor is proportional to the extrinsic curvature (peculiar velocities are not small but vanish), we have vanishing shear 
everywhere and this implies homogeneity in cosmologically relevant cases\footnote{Note that {\it a priori} the tilt of the $4-$velocity relative to the hypersurface normal is not specified. For
vanishing tilt, as considered here, and in the case of dust matter, shear--free motion implies homogeneity; this also holds true for
large classes of perfect fluid models, see \cite{colwai83,collinswhite,collins88}. 
We briefly show for the case of dust matter that we can determine the lapse function such that the model is {\it hypersurface--homogeneous}: we use the momentum constraints, $K^i_{\;j || i} - K^k_{\;k | j} =0$, for the extrinsic curvature of Eq.~(\ref{shearfree}), and integrate them to yield $N s(t) =  \psi' / (1-2\psi )- \tilde H $,
with a time--dependent function of integration $s(t)$ that reflects the
freedom of time--reparametrization. On the other hand we have from $\dot{\cal S} = N^{-1}{\cal S}_{,t} = h$, with $h = (\varepsilon+p)/\varrho$:
$N = {\cal S}_{,t} / h$. Equating the two relations
for the lapse function gives $ \psi' / (1-2\psi )- a' / a = s(t) {\cal S} (t) / h$.
This shows that, if we require $p=0$ and hence $N_{|i} =0$ and $h=1$, the function $\psi$ can only be time--dependent and does
not describe perturbations in the considered hypersurfaces; e.g. the Ricci tensor, given further below in explicit form, vanishes everywhere, since the space gradients of $\psi (t)$
vanish. For further discussions of this metric form including estimates, see \cite{estim}. Work is in progress that analyzes the present issue in a tilted slicing and non--vanishing shift vector.}.

In the sequel we shall consider (\ref{postnewtonlapseandmetric}) as an {\it ansatz} for the
metric and only later, in the final result, we may look at the importance of the various terms; 
we do not invoke any further approximation in the following calculations. The usual practice of 
employing perturbative assumptions from the outset may mask the simplicity of the problem.
Also, we prefer to not specify the lapse function; we shall retain $N$
so that we can discuss the result for different foliations. 
We introduce, to shorten the notation, the auxiliary variable with its derivatives with respect to the coordinate time:
\begin{equation}
\label{morphon1}
\fl\qquad
\alpha :=-\ln\sqrt{1-2\psi} \;\;\;;\;\;\;\alpha' = \frac{\psi'}{1-2\psi}\;\;\;;\;\;\;
\alpha'' = \frac{\psi''}{1-2\psi} + 2\frac{(\psi' )^2}{(1-2\psi)^2}\;\;.
\end{equation}

\subsubsection*{Kinematics of the volume}

Let us start with the simple observation that, given the $3-$metric in the form 
(\ref{postnewtonlapseandmetric}), the kinematics of the volume is determined by
a given solution for $\psi$:
the volume of an averaging domain, and hence the effective scale--factor $a_{\cal D}$, 
is calculated from its definition $V_{\cal D} = \int_{\cal D}J d^3 X$ with 
$J= \sqrt{\det (g_{ij})} = a^3 (1-2\psi)^{3/2}$:
\begin{equation}
\label{volume}
V_{\cal D} =  a_{\cal D}^3 {V_{\cal D}}_i =
\int_{\cal D} a^3 \left[\,1-2\psi \,\right]^{3/2} d^3 X  \;.
\end{equation}
The {\it rate of volume expansion} of a domain in the spatial hypersurface, written in terms
of the Hubble functional $H_{\cal D}(t)$, reads:
\begin{equation}
\label{volumeexpansion}
\frac{1}{3}\langle N\theta\rangle_{\cal D} = \frac{1}{3}\frac{V_{\cal D}'}{V_{\cal D}}=
{H}_{\cal D} = \langle{\tilde H}\rangle_\CD - \langle \alpha' \rangle_{\cal D}\;\;,
\end{equation}
with the local Hubble function $\tilde H (X^i ,t): = a' /a$ of the background model as seen in the spatial hypersurfaces specified by the
inhomogeneous lapse, i.e. ${H}= {\dot a}/a = a' /(a N ) = \tilde{H}/N$.
At this place note that there is no ambiguity concerning the notion of {\it averaged volume
expansion}, once the lapse function has specified the foliation of spacetime.
Although being unambiguous, we have to come back to this point later, since a majority
of papers on averaging scalar metric inhomogeneities employs Euclidean volume averaging
on the background metric in a frame of global coordinates, which do not exist in a general
relativistic setting, if the space is deformed by inhomogeneous perturbations (see the related discussion of George Ellis \cite{ellisFOCUS}).

The second time--derivative provides the kinematical equation for the 
{\it rate of volume acceleration}: writing
\begin{equation} 
\frac{a_{\cal D}''}{a_{\cal D}} = H_{\cal D}' + H_{\cal D}^2 \;,
\end{equation}
and, using the commutation rule (\ref{commutation}) for both terms on the r.--h.--s., we calculate: $H_{\cal D}' =$
\begin{equation*}
\fl 
\langle{\tilde H}'\rangle_{\cal D} -\langle \alpha''\rangle_{\cal D}
+ 3[\langle(\alpha' )^2 \rangle_{\cal D} - \langle\alpha' \rangle_{\cal D}^2 ]
- 6[\langle {\tilde H}\alpha' \rangle_{\cal D} - \langle{\tilde H}\rangle_{\cal D}
\langle \alpha' \rangle_{\cal D}]
+ 3[\langle{\tilde H}^2 \rangle_{\cal D} - \langle{\tilde H}\rangle_{\cal D}^2 ],\;{\rm and}\nonumber
\end{equation*}
$H_{\cal D}^2 = \langle \alpha'\rangle_{\cal D}^2 -2 \langle{\tilde H}\rangle_{\cal D}
\langle\alpha' \rangle_{\cal D} + \langle{\tilde H}\rangle_{\cal D}^2$.
Summing up the above terms, using ${\tilde H}' + {\tilde H}^2 = a'' / a$,
and rearranging some terms, we get for the rate of volume acceleration:
\begin{eqnarray}
\label{volumeacceleration}
 \frac{a_{\cal D}''}{a_{\cal D}} =\Bigl\langle \frac{a''}{a}\Bigr\rangle_{\cal D}
+ 2[\,\langle(\alpha' )^2 \rangle_{\cal D} - \langle\alpha' \rangle_{\cal D}^2 \,]
\qquad\nonumber\\
- 4[\,\langle {\tilde H}\alpha' \rangle_{\cal D} - \langle{\tilde H}\rangle_{\cal D}
\langle \alpha' \rangle_{\cal D}\,]
+ 2[\,\langle{\tilde H}^2 \rangle_{\cal D} - \langle{\tilde H}\rangle_{\cal D}^2 \,]\nonumber\\
- \langle\alpha'' + 2{\tilde H}\alpha' - (\alpha' )^2 \rangle_{\cal D}\;\;,
\qquad\qquad
\end{eqnarray}
where $a'' /a = N^2 (\ddot a / a ) + N' (\dot a / a)$ is the local rate of 
acceleration of the background model as seen in the spatial hypersurfaces.

This latter equation is quite simple given the fact that the second line above only
appears, since we have factored out a ``background model'' which, within the
spatial hypersurfaces, appears as a fluctuating background:
there is a fluctuating $t-$Hubble function, which with the replacement 
${\tilde H}= N H$ results in a {\it frame backreaction} term. 
This term has to be taken seriously, since there are no ``background observers'' from
which we could see a homogeneous behavior of the background expansion.

Given a solution of Einstein's equations for $\psi$ and given a  lapse, e.g., as a 
functional of $\psi$, the above equation provides the 
answer to the question, whether the universe model `accelerates'. 
We did not use any dynamical equations of general relativity so far.

\subsubsection*{Backreaction terms}

Since we did not employ any approximation beyond those of the chosen framework, 
the first and second $t-$derivatives of the scale factor calculated above are in accord with
those found from the general equations (\ref{effectiveequations}). To demonstrate this we have to employ 
dynamical equations of general relativity. We evaluate the  
kinematical and dynamical backreaction terms:
\begin{equation}
\label{Q}
\fl\qquad
\frac{{\tilde{\cal Q}}_{\cal D}}{6} =   
[\,\langle(\alpha' )^2 \rangle_{\cal D} - \langle\alpha' \rangle_{\cal D}^2 \,]
- 2 [\,\langle {\tilde H}\alpha' \rangle_{\cal D} - \langle{\tilde H}\rangle_{\cal D}
\langle  \alpha' \rangle_{\cal D}\,]
+ [\,\langle{\tilde H}^2 \rangle_{\cal D} - \langle{\tilde H}\rangle_{\cal D}^2 \,]\;;
\end{equation}
\begin{equation}
\label{P}
\fl\qquad
{\tilde{\cal P}}_{\cal D} = \langle {\tilde{\cal A}}\rangle_{\cal D}
+ 3 \Bigl\langle \frac{N'}{N} \left(\,{\tilde H} - \alpha' \,\right)\Bigr\rangle_{\cal D}\;\;.
\end{equation}
We employ the Hamiltonian constraint and Raychaudhuri's equation in the forms:
\begin{eqnarray}
\label{ham}
{\tilde {\cal R}}=16\pi G {\tilde\varepsilon} + 2\Lambda N^2 - 6 {\tilde H}^2 + 12 {\tilde H}\alpha'
-6 (\alpha' )^2 \;\; \\
\label{ray}
\alpha'' + 2{\tilde H} \alpha' - (\alpha' )^2 
= \frac{4\pi G}{3}\left(\,{\tilde\varepsilon} + 3{\tilde p}\,\right)
- \frac{N'}{N} ({\tilde H} - \alpha' ) - \frac{1}{3} {\tilde{\cal A}} + \frac{a''}{a}\;.
\end{eqnarray}
Then, we take  spatial averages, use the averaged $t-$scalar curvature $\langle N^2 {\cal R}
\rangle_{\cal D}$, and insert the average of 
the left--hand--side of (\ref{ray})  into the last line of Eq.~(\ref{volumeacceleration}); 
we so find consistency with the averaged equations (\ref{effectiveequations}):
\begin{equation}
\fl\quad
\frac{a_{\cal D}''}{a_{\cal D}} = - \frac{4\pi G}{3}\langle{\tilde\varepsilon} +
3{\tilde p}\rangle_{\cal D} + \frac{1}{3}\left(\, {\tilde{\cal Q}}_{\cal D}+ 
{\tilde{\cal P}}_{\cal D}\, \right)\;\;;\;\;
6H_{\cal D}^2 = 16\pi G \langle{\tilde\varepsilon}
\rangle_{\cal D} - {\tilde{\cal Q}}_{\cal D}
- \langle{\tilde{\cal R}}\rangle_{\cal D}\;.
\end{equation}
(The $t-$averaged curvature term cancels in the averaged Raychaudhuri equation.)

\subsubsection{Dynamical equations for scalar metric perturbations}

The Hamiltonian constraint and Raychaudhuri's equation may be rewritten by
replacing the coordinate time--derivative (denoted by a prime in the previous equations) 
through the covariant derivative (denoted by an overdot),
and use of the relation ${\tilde H}= H N$: 
\begin{eqnarray}
\label{constraintscov}
{\cal R} = 16\pi G \varepsilon + 2 \Lambda - 6 H^2 + 12 {H}
{\dot\alpha}  - 6 {\dot\alpha}^2 \;\;;\\
\label{raycov}
{\ddot\alpha} + 2H {\dot\alpha} - {\dot\alpha}^2 -\frac{4\pi G}{3}\left(\,\delta\varepsilon + 3\delta p\,\right)
+\frac{1}{3}{\cal A}=0\;\;,
\end{eqnarray}
where we have split off the background model through introduction of the deviations  $\delta{\varepsilon} := {\varepsilon}- {\varepsilon}_H$ and
$\delta{p} := {p} - {p}_H$, and where the background quantities have to obey the covariant equation
\begin{equation}
\frac{\ddot a}{a} + \frac{4\pi G}{3} ({\varepsilon}_H + 3  p_H)=0\;\;\;\Leftrightarrow\;\;\;\frac{a''}{a} + \frac{4\pi G}{3}({\tilde\varepsilon}_H + 3 {\tilde p}_H) = \frac{N'}{N}{\tilde H}\;.
\end{equation}
Equation~(\ref{raycov}) is key to the determination of solutions for $\psi$; we therefore write it in terms of $\psi$:
\begin{equation}
\label{raycovpsi}
{\ddot\psi} + 2H{\dot\psi} -\frac{1}{3} \left(\,4\pi G [\,\delta\varepsilon + 3\delta p \,] - {\cal A}\,
\right)(1-2\psi)+ \frac{{\dot\psi}^2}{1-2\psi}=0 \;\;. 
\end{equation}
The acceleration divergence ${\cal A}= {\tilde{\cal A}}/N^2$ in these equations can be explicitly expressed through 
$\psi$ only, if the lapse function is specified, e.g. through $\psi$. If it is specified according to (\ref{postnewtonlapseandmetric}) we obtain:
\begin{equation}
\label{accelerationdivergence}
{\cal A} = \frac{N^{|i}_{\;\,||i}}{N} = \frac{\phi^{|i}_{\;\,||i}}{1+2\phi} -  \frac{\phi^{|i}\phi_{|i}}{(1+2\phi)^2}\;\;,
\end{equation}
involving the Laplace--Beltrami operator $\Delta_g$ on the 3--metric,
(note: $\Gamma^i_{\;\,ik} = (\ln J)_{|k}$ and $J := \sqrt{\det (g_{ij})} = a^3 (1-2\psi )^{3/2}$):
\begin{eqnarray}
\label{laplacebeltrami}
\Delta_g \phi :=\phi^{|i}_{\;\,||i} = [\,g^{ij}\phi_{|j}\,]_{||i}=[\,g^{ij}\phi_{|j}\,]_{|i}+ g^{kj}\phi_{|j}
\Gamma^i_{\;\,ik} 
= \frac{1}{J}[\,g^{ij}\,J\,\phi_{|j}\,]_{|i}\;.
\end{eqnarray}
We notice that we have to specify the form of the lapse function. Then, we can proceed 
by solving or approximating the local evolution equation
governing $\psi$. Also this problem, if approached perturbatively from the outset, could mask
the transparency of the general problem, and so may give rise to ambiguities in interpretation.

\subsubsection*{Example}

As an example that connects the above thoughts with the model in Subection~\ref{subsect:RZA} we shall specify the matter model to an {\it irrotational dust continuum},
i.e. $p=0$, and therefore $h=1$, and $\varepsilon = \varrho$. We consider the comoving synchronous gauge with $N=1$ without loss of generality.
Hence, we have no acceleration in the hypersurfaces, 
${\cal A}=0$, and retain the simplified equation:
\begin{equation}
\label{dustevolution}
{\ddot\psi} + 2H{\dot\psi} -\frac{4\pi G}{3} \delta\varrho (1-2\psi)+ \frac{{\dot\psi}^2}{1-2\psi}=0\;\;;\;\; \delta\varrho = \varrho - \varrho_H = : \varrho_H \delta \;\;,
\end{equation}
where the overdot is, in this gauge, equivalent to the coordinate time--derivative, and $\delta$ is defined as the conventional density contrast.
Since the energy density is reduced to the restmass density $\varrho$, we can employ
the general integral 
\begin{equation}
\label{densityintegral}
\varrho = \varrho_{initial} J^{-1}\;\;\;{\rm i.e.}\;\;\;
\varrho = \frac{1}{a^3}\frac{\varrho_{initial}}{(1-2\psi )^{3/2}} = \varrho_H \frac{1 + \delta_{initial}}{(1- 2 \psi )^{3/2}}\;\;.
\end{equation}
Although a general solution to Eq.~(\ref{dustevolution}) may exist and could be found, we here
resort to approximations. Linearizing the above equation 
with respect to $\psi$, i.e. also expanding $\delta\varrho (1- 2\psi )$
to linear order, $\delta\varrho (1 - 2 \psi ) \approx \varrho_H [ (1 +\delta_{initial}) (1+3\psi) - 1 ] ( 1 - 2\psi ) \approx \varrho_H (\delta_{initial} + 3\psi)$, we obtain a familiar
equation for the approximate solution $\psi^A$ \cite{gaugeinv}:
\begin{equation}
\label{dustevolutionlinear}
{\ddot\psi}^A + 2H{\dot\psi}^A -4\pi G \varrho_H \psi^A = \frac{4\pi G\varrho_H \delta_{initial}}{3}\;\;.
\end{equation}
Interestingly, this approximation may be interpreted as the scalar part of the (first--order) relativistic 
Lagrangian approximation discussed in Subsection~\ref{subsect:RZA}. For, if we write the $3$--metric in terms of Cartan one--forms,
$g_{ij} = \delta_{ab}\eta^a_{\;i}\eta^b_{\;j} = a^2 (1-2\psi)\delta_{ij}$, 
the deformation is given by $\eta^a_{\;i} = a \sqrt{1-2\psi}\,\delta^a_{\;i}$.
On the other hand, the equation for the trace of first--order Lagrangian perturbations reads \cite{RZA1}: 
\begin{equation}
\label{dustdeformationlinear}
\fl
^A{\ddot{\eta^a_{\;\,i}}}= a [ \delta^a_{\;i} + P^a_{\;\,i} ]\;\;;\;\;P: = P^k_{\;\,k} = \delta_{ak}P^a_{\;\,k}\;\;;\;\;
\ddot P + 2H P -4\pi G \varrho_H P = 4\pi G\varrho_H \delta_{initial}.
\end{equation}
Linearizing the trace of the Cartan deformation above we obtain the relation $P = 3\psi^A$, so that 
Equation~(\ref{dustdeformationlinear}) implies Equation~(\ref{dustevolutionlinear}).
Keeping only the growing part of the homogeneous solution we have:
\begin{equation}
\label{solution}
\psi^A = \psi_{initial} (X^i) \xi (t)\;\;\;{\rm with}\;\;\;{\ddot \xi} +2H {\dot \xi} - 4\pi G \varrho_H (\xi + 1) = 0\;\;, 
\end{equation}
with the well--known solutions $\xi (t)$ of the linear perturbation theory, or the standard Zel'dovich approximation \cite{zeldovich,buchert89,buchert92,bbk}.

Inserting this approximate solution into the general averaged equations results in a non--perturbative backreaction model.
The averaged intrinsic curvature can be
evaluated in two ways: firstly, from averaging the Hamiltonian constraint (\ref{ham})
as above (which is enough for the kinematically averaged equations),
and, secondly, from the geometry of the spatial hypersurface: locally, we have for the
$3-$Ricci tensor:
\begin{equation}
\label{riccipsi}
{\cal R}^i_{\;\,j}= \frac{1}{a^2}\left[\,\frac{\delta^{ik}\psi_{|kj} + 
\Delta\psi \delta^i_{\;j}}{(1-2\psi)^2} + \frac{3\delta^{ik}\psi_{|k}\psi_{|j} + 
(\nabla\psi)^2\delta^i_{\;j}}{(1-2\psi)^3}\,\right]\;,
\end{equation}
and for the scalar $3-$curvature ${\cal R}=  {\cal R}^k_{\;\,k}$:
\begin{equation}
\label{Rpsi}
{\cal R}= \frac{4}{a^2}\left[\,\frac{\Delta\psi}{(1-2\psi)^2} + \frac{3}{2}
\frac{(\nabla\psi)^2}{(1-2\psi)^3}\,\right]\;\;,
\end{equation}
with the Euclidean operators $\Delta\psi = \delta^{ik}\psi_{|i k}$,
$\nabla\psi = \psi_{|k}$. (Note that, since $N=1$, the scale--factor $a$ depends only on the coordinate--time $t$; for non--constant lapse the above expression also contains spatial derivatives of this scale--factor.) Averaging the above curvature expression can be used to test
consistency and therefore the quality of the proposed non--perturbative approximation for the backreaction model.

\subsubsection*{Discussion: non--perturbative versus perturbative approaches\\}

The above example calculation for metrical inhomogeneities of dust furnishes a perturbative approach with regard to the local
solution. Considering the backreaction problem, i.e. the evolution of average properties given the local solution,
we are entitled to look at the averaged equations in their general form and only approximate the local
evolution of $\psi$, and this as general as we can (compare here some results obtained in the Newtonian theory
\cite{buchert:nonperturbative}). In such an approach no approximation is done on functionals
of $\psi$ like the averaged equations, the averaging operator itself, etc.
We so deal with a non--perturbative result. The perturbative result should arise as a limit; however, this limit process
is drastic: it involves not only linearizing the equations, but also the averaging operator, which
explains that in previous work many more backreaction terms were found \cite{abramo1}; compare also \cite{clarkson:perturbations,umeh} and the explicit discussion of this point in \cite{chrisreview} and the contribution \cite{chrisFOCUS} to this volume.
Exploiting our result in this limit is of course possible, but from a physical point of view not necessary. The present suggestion of 
a non--perturbative approach consists in performing approximations only for the fluctuating local sources, but how these approximations enter functional 
expressions like the volume average and the backreaction model is not further approximated.  

The discussion of scalar metric inhomogeneities also should make clear that for a realistic modeling of inhomogeneities the tensorial degrees of freedom have to be taken into account. As our example and its connection to the relativistic Zel'dovich approximation show, this can be achieved and forms the subject of forthcoming investigations \cite{RZA1,RZA2}.

\subsubsection*{Acknowledgements:} {\footnotesize
Thanks go to my collaborators with whom I share some of the presented results, especially to Henk van Elst and Xavier Roy for valuable comments on the manuscript. 
This review is an extension of lecture notes \cite{mexico} (Sections 1--6), and
two papers in preparation (Sections 7.1 -- 7.3), as well as a draft written in 2005 (Section 7.4). It is a pleasure to thank Lars Andersson and Alan Coley for inviting this contribution.
This work is supported by ``F\'ed\'eration de Physique Andr\'e--Marie Amp\`ere'' of Universit\'e Lyon 1 and \'Ecole Normale Sup\'erieure de Lyon. }

\section*{References ({\em blue links refer to articles in the Focus Section})}
\end{document}